\documentclass[prb,showpacs,twocolumn,superbib,floats,,superscriptaddress]{revtex4-1}

\usepackage{graphicx}
\usepackage{amsfonts}
\usepackage{amsmath}
\usepackage{natbib}
\usepackage{multirow}
\usepackage{bm}
\usepackage[dvips]{epsfig}
\usepackage{color}
\usepackage{ulem}
\usepackage{caption}
\usepackage{floatrow}
\usepackage{braket}

\begin{document}

\title{Temperature dependent excitonic effects in the optical properties of single-layer MoS$_2$}

\author{Alejandro Molina-S\'{a}nchez}
\affiliation{Physics and Materials Science Research Unit, University of Luxembourg, 162a avenue de la Fa\"iencerie, L-1511 Luxembourg, Luxembourg}

\author{Maurizia Palummo}
\affiliation{University of Rome Tor Vergata, Rome, Italy}
\affiliation{INFN, Laboratori Nazionali di Frascati, Via E. Fermi 40, 00044
Frascati, Italy}

\author{Andrea Marini}
\affiliation{Istituto di Struttura della Materia of the National Research Council, Via Salaria Km 29.3, I-00016 Monterotondo Stazione, Italy}

\author{Ludger Wirtz}
\affiliation{Physics and Materials Science Research Unit, University of Luxembourg, 162a avenue de la Fa\"iencerie, L-1511 Luxembourg, Luxembourg}

\date{\today}

\begin{abstract}
Temperature influences the performance of 
two-dimensional materials in optoelectronic devices. Indeed, the optical
characterization of these materials is usually realized at room temperature.
Nevertheless most {\it ab-initio} studies are yet performed without including any
temperature effect. As a consequence, important features are thus overlooked,
such as the relative intensity of the excitonic peaks and their broadening,
directly related to the temperature and to the non-radiative exciton relaxation
time.
We present {\it ab-initio} calculations of the optical response of single-layer MoS$_2$, a prototype 2D material,
as a function of temperature using density functional theory and 
many-body perturbation theory. We compute the electron-phonon interaction using
the full spinorial wave functions, i.e., fully taking into account effects of spin-orbit interaction. We find that bound excitons ($A$ and $B$ peaks) and
resonant excitons ($C$ peak) exhibit different behavior with temperature,
displaying different non-radiative linewidths. 
We conclude that the inhomogeneous broadening of the absorption
spectra is mainly due to electron-phonon scattering mechanisms. Our calculations
explain the shortcomings of previous (zero-temperature) theoretical spectra
and match well with the experimental spectra acquired at room temperature. 
Moreover, we disentangle the contributions of acoustic and
optical phonon modes to the quasi-particles and exciton linewidths. Our model also allows to identify which phonon modes couple to each exciton state, useful for the interpretation of resonant Raman scattering experiments.
\end{abstract}


\maketitle

\section{Introduction}

Ultrathin two-dimensional materials such as graphene and MoS$_2$ are appealing candidates
for a new generation of opto-electronic devices\cite{geim_rise_2007} such as photoresponsive memories\cite{roy_graphene-mos2_2013}, light-emitting and harvesting devices\cite{lopez-sanchez_light_2014}, or 
nano-scale transistors\cite{radisavljevic_single-layer_2011}. 
They are also suitable platforms for carrying out research
on fundamental physics phenomena like the valley Hall 
effect,\cite{mak_valley_2014} ultrafast charge 
transfer,\cite{hong_ultrafast_2014} or valley excitons in two-dimensional materials.\cite{yu_valley_2015}  Technologically, single-layer MoS$_2$ 
is relevant due to a direct optical gap at 1.8 eV and a high electron mobility.\cite{lembke_breakdown_2012}

The optical response of MoS$_2$ is dominated by strongly bound 
excitons.\cite{Komsa2012,molina-sanchez_effect_2013,qiu_optical_2013} 
The same holds for the other group VI
semiconducting single-layer transition-metal dichalcogenides (TMDs) such 
MX$_2$ with M = Mo or W and X = S, Se, or Te\cite{Komsa2013,surfscirep_alejandro}. This suggests their possible use in opto-electronic devices working 
at room temperature. Nevertheless most of the 
modern first-principles ground and excited state simulations are performed 
at 0 K and thus omit the role of thermal lattice vibrations 
on the electronic and optical properties.   

In general, temperature has a capital influence on the electronic and optical properties of  
semiconductors determining their application as optoelectronic devices.\cite{wolpert_temperature_2012} 
It is well known that its drives the  
band gap renormalization,\cite{lautenschlager_temperature_1985} and induces changes in 
the position and width of the optical peaks.\cite{lautenschlager_temperature_1987} 
At the same time also the spectra obtained from
other techniques such as angle-resolved photoemission spectroscopy (ARPES) \cite{Miwa2015} 
are clearly influenced by the temperature due to 
the enhanced mixing of electron and phonon states. 

The possibility to perform electronic structure calculations based on {\it ab-initio} approaches including 
the electron-phonon (EP) interaction is thus of paramount importance.
Even though many years ago \cite{cardona_electronphonon_2005} Heine, Allen and Cardona (HAC) pointed out that the EP coupling can induce corrections
of the electronic levels as large as those induced by the electronic correlation, 
the number of works based on first-principles simulations addressing this problem is still very limited
and mainly dedicated to traditional bulk compounds.\cite{giustino_electron-phonon_2010} 
The inclusion of EP couplings considerably broadens the scope of 
first-principles electronic-structure calculations beyond the study of
temperature effects. It opens the way to the study of 
many interesting phenomena such as polaron formation in crystals and transport properties.\cite{moser_tunable_2013} 

As for MoS$_2$, theoretical {\it ab-initio} studies including the EP 
interaction have addressed  specific aspects such as 
phonon-limited mobility,\cite{kaasbjerg_phonon-limited_2012} thermal
conductivity,\cite{li_thermal_2013} electron
cooling,\cite{kaasbjerg_hot-electron_2014} or electron
transport.\cite{ge_effect_2014} Tongay \textit{et. al.}\cite{Tongay2012}
have performed a theoretical and 
experimental study of the band gap dependence on
temperature for multi-layer MoSe$_2$ and MoS$_2$. They have attributed
all temperature effects to lattice renormalization induced by the thermal
expansion. They capture
correctly the band gap trend only for high temperatures (above 300 K). Below
room temperature, the electron-phonon interaction plays a crucial role but it is
ignored by Tongay \textit{et. al.}\cite{Tongay2012} In Ref. \onlinecite{qiu_optical_2013} Qiu
\textit{et. al.} have studied temperature effects by including
the quasi-particle linewidths. However, they ignored the energy renormalization and the
accurate calculation of the linewidths across all the Brillouin zone. Here we
explore using a fully \textit{ab initio} approach
how the EP interaction induces changes in the electronic structure
and optical properties of the MoS$_2$ single-layer. This also enables us to address
photoluminescence,\cite{li_measurement_2014} 
ARPES,\cite{latzke_electronic_2015,Miwa2015} and resonant Raman
scattering experiments.\cite{carvalho_symmetry-dependent_2015} 

Differently from most of the recent works on bulk materials
\cite{antonius_many-body_2014}
we do not limit our study to the band gap renormalization but we 
extend our investigation to the full band structure, with special attention to 
the electron states of interest for opto-electronic applications. 
Starting from previous studies which established the existence
of several kind of excitonic 
states in this low-dimensional
material,\cite{molina-sanchez_effect_2013,qiu_optical_2013} whose behavior we
characterize as a function of the temperature. We calculate the shift of the binding energy and the 
non-radiative linewidths of excited states.

It is worth to underline that in
our approach we use the full  
spinorial nature of the wave functions through all ground and excited state calculations. 
This is quite important because it is well known that spin-orbit coupling determines the valley polarization dynamics and is  fundamental 
to understand the optical properties of all TMDs.  

\section{The theoretical approach}

Our calculations start with density-functional theory (DFT) to obtain a first estimate of the electronic bands. We use density-functional perturbation theory (DFPT) to calculate the phonon modes and the electron-phonon coupling matrix
elements. With the latter we calculate the change of the electronic bands  due 
to the lattice vibrations.\cite{Marini2015} Afterwards, we solve
the temperature dependent Bethe-Salpeter equation\cite{marini_textitab_2008}. 
We thus
explore the change in the optical spectra, and in the exciton energies
and linewidths when temperature increases.

The ground state properties of single-layer MoS$_2$, eigenvalues and
wave functions, are calculated with the QUANTUM ESPRESSO 
code\cite{giannozzi_quantum_2009} within the local density approximation (LDA) for the exchange-correlation potential.
We use DFPT to obtain the phonon modes as well as the first and second 
order electron-phonon matrix elements.\cite{cannuccia_effect_2011} As mentioned in the introduction the spin-orbit interaction is essential
to correctly describe
excitons in MoS$_2$, for this reason also the 
electron-phonon matrix elements are calculated 
taking into account the full spinor wave functions. 

We study the temperature effects on the electronic states and on the excitons 
by merging DFT/DFPT with many-body perturbation theory. 
Within this framework, two self-energy  
diagrams, which correspond to the lowest non vanishing terms of a perturbative 
treatment, have to be evaluated.
The Fan self-energy,\cite{cannuccia_effect_2011} related to first order terms

\begin{multline}
\Sigma^{Fan}_{n,\bf k}(\omega,T) = i\sum_{n' \bf{q}\lambda} \frac{|g^{\bf{q}\lambda}_{nn' \bf{k}}|}{N_q}\times \\ \times\left[ \frac{N_{\bf{q}}(T)+1-f_{n'\bf{k-q}} }{\omega-\varepsilon_{n'\bf{k-q}} - \omega_{\bf{q}\lambda} - i0^{+} } \right] \\ \times
\left[ \frac{N_{\bf{q}}(T)+f_{n'\bf{k-q}} }{\omega-\varepsilon_{n'\bf{k-q}} + \omega_{\bf{q}\lambda} - i0^{+} } \right] ,
\label{eq_fan}
\end{multline}

where $\varepsilon_{n,\bf k}$ are the LDA eigenvalues, $\omega_{\bf{q},\lambda}$ the 
phonon frequencies, $f_{n,\bf k}$ and $N_{\bf q}(T)$ are the Fermi and Bose distribution
functions of electrons and phonons, respectively. The self-energy associated
to an electron state $(n,\bf{k})$ is the sum over
all the electron states $n'$ and phonon modes $\lambda$, 
where $N_q$ is the number of $\bf q$ vectors in the Brillouin zone.
Conservation of momentum is explicitly enforced. The first order electron-phonon
matrix elements $g_{nn'\bf k}^{\bf q\lambda}$ represents the amplitude for the scattering  process
$|n {\bf k}\rangle \rightarrow |n' {\bf k-q}
\rangle\otimes |{\bf q}\lambda \rangle$\cite{Marini2015}. We have a similar expression for the
Debye-Waller (DW) self-energy, related to the second order terms,

\begin{equation}
\Sigma^{DW}_{n,\bf k}(T) =\frac{1}{N_q} \sum_{\bf{q}\lambda} \Lambda_{nn\bf
k}^{\bf q \lambda, \bf -q \lambda}(2N_{\bf{q}\lambda}(T)+1),
\label{eq_dw}
\end{equation}

where $\Lambda_{nn'\bf k}^{\bf q \lambda, \bf q' \lambda'}$, insted, represents the amplitude for the second--order scattering process
$|n {\bf k}\rangle \rightarrow |n',{\bf k-q -q'}\rangle  \otimes |{\bf q}\lambda\rangle \otimes |{\bf q}' \lambda' \rangle $\cite{Marini2015}.
In both self-energy terms, temperature enters via the phonon population. In
polar semiconductors the electron-phonon interaction strength becomes larger
when including the Fr\"{o}hlich polar-coupling term.\cite{Verdi2015} Nonetheless,
the LO-TO splitting in MoS$_2$ is rather small, 3 cm$^{-1}$, and we do not expect
significant changes in single-layer.

The fully interacting electron propagator (accounting for the 
electron-phonon interaction) is

\begin{equation}
G_{n\bf{k}}(\omega,T)=\left(\omega - \varepsilon_{n\bf{k}} - 
\Sigma^{Fan}_{n\bf{k}}(\omega,T) - 
\Sigma^{DW}_{n\bf{k}}(T) \right)^{-1}.
\end{equation}

The complex poles of this equation define the electronic excitations
of the interacting system. If the quasi-particle approximation (QPA) is valid, 
and assuming a smooth frequency dependence, the electron-phonon self-energy can be expanded up to
the first order around the bare energies ($\epsilon_{nk}$). 
In this case the temperature dependent quasi-particle energies are defined  
as, \cite{marini_yambo:_2009}

\begin{equation}
E_{n\bf{k}}(T) = \varepsilon_{n\bf{k}} + Z_{n\bf{k}}(T)\left[ \Sigma^{Fan}_{n\bf{k}}(\varepsilon_{n\bf{k}},T) + \Sigma^{DW}_{n\bf{k}}(T)\right].
\label{qp-energy}
\end{equation}

It is clear that the quasi-particle energies 
depend on temperature and are complex numbers, where  the real parts
are the quasi-particle energies and the imaginary parts, $\Gamma_{n\bf{k}}(T)$, correspond 
to the quasi-particle widths. The 
renormalization factor $Z_{n\bf{k}}$ represents the quasi-particle
charge. Therefore the QPA makes sense when $Z_{n\bf{k}}$ takes a value close to 1. 
If the QPA holds, the spectral function,
$A_{n\bf{k}}(\omega,T)$, the imaginary part of the Green's function,
is a single peak Lorentzian function centred at $E_{n\bf{k}}$ and with
width $\Gamma_{n\bf{k}}(T)$. The narrower the spectral function is the weaker 
the electron-lattice interaction is. 
When the electron-lattice interaction becomes strong
enough, the QPA breaks down, the spectral functions does no longer consist
of Lorentzian peaks but span a wide energy range.\cite{cannuccia_effect_2011}

\section{Temperature dependent electronic structure of single-layer MoS$_2$}

The calculation of the electronic structure of MoS$_2$ has been done in a plane-wave 
basis using norm-conserving pseudopotentials and a kinetic energy cutoff of 80 Ry and a $\bf k$-grid
of $12\times 12\times 1$. On top of self-consistent DFT simulations, electron-phonon matrix elements are obtained by DFPT in the local-density approximation.
From the explicit expressions of the self-energy terms (Eqs.~(\ref{eq_fan}) and (\ref{eq_dw})), it becomes clear that a careful 
convergence over the number of bands $n'$ 
and the number of transferred phonon momenta $\bf{q}$ to evaluate the integral over the Brillouin zone,
is required. From our study we have found that, the spectral
functions of MoS$_2$ converge using a set of 400 randomly distributed $\bf{q}$
points and 
36 bands (18 occupied bands - we do not take into account Mo-semicore electrons - 
and 18 empty bands). The calculations are converged with respect to number of
$\bf{q}$-points and bands. We have checked this on the profile of the spectral function
which is a more stringent test than checking the eigenvalue correction. We have used a Lorentzian broadening of 60 meV.
Recent works on diamond  
and silicon required a much larger number of bands and $\bf q$-points
to reach convergence.\cite{Ponce2015}. The rapid convergence with the number
of $\bf q$-points we have found 
here is mainly due to the two-dimensional nature of the material under
investigation.

\begin{figure*}
\includegraphics[width=7    cm]{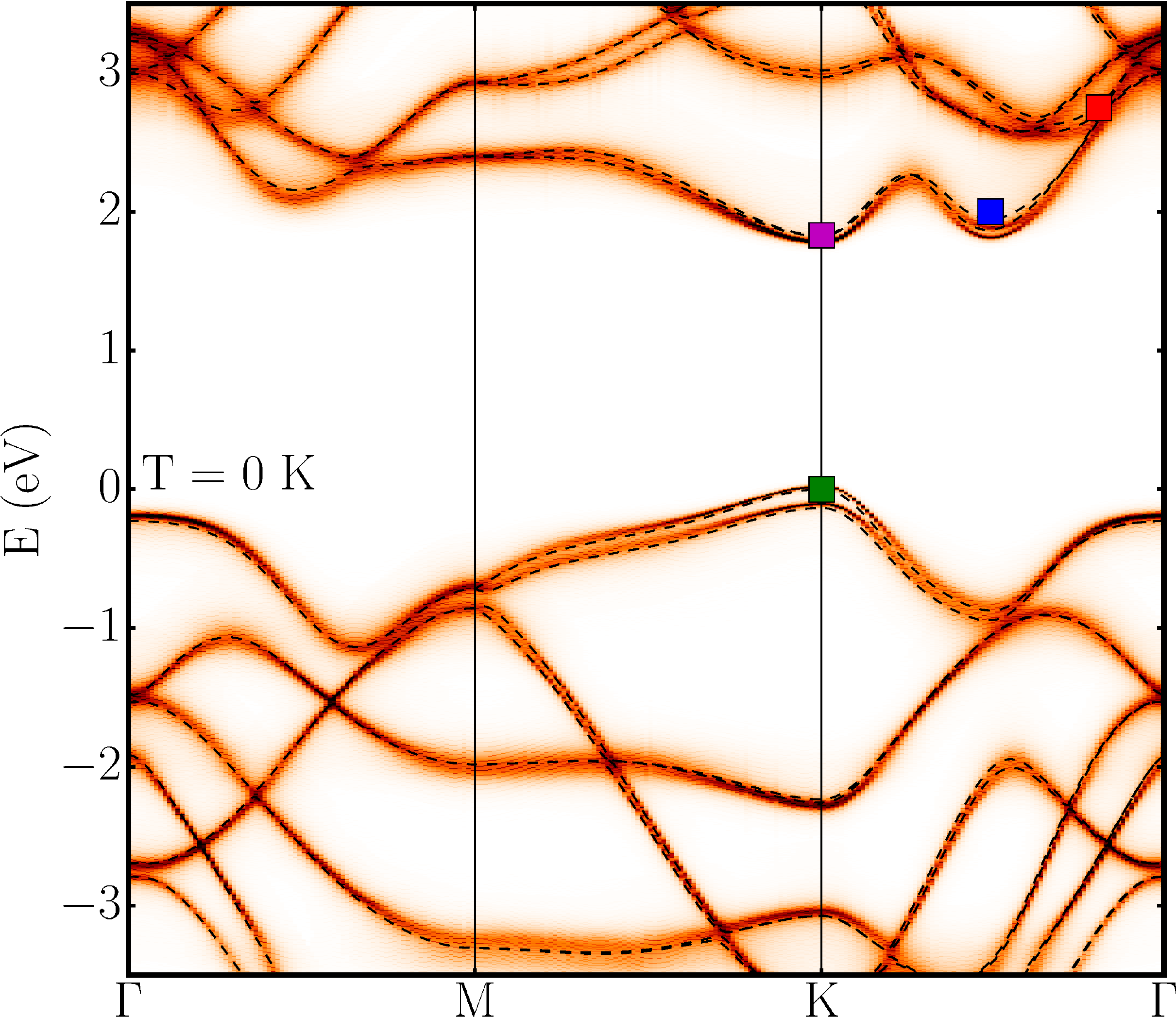}
\includegraphics[width=7.05 cm]{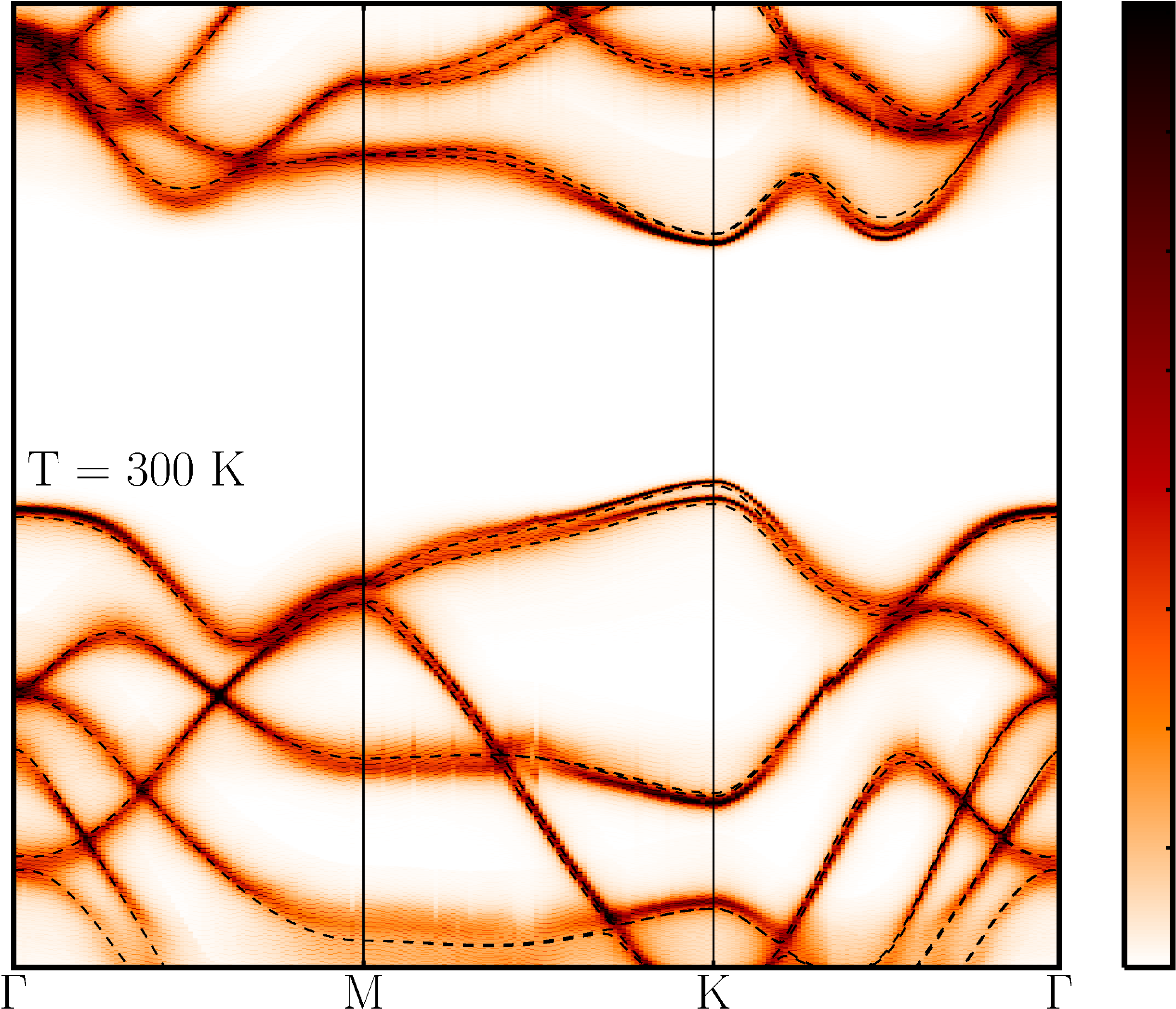}
\caption{Spectral functions of single-layer MoS$_2$ 
for temperatures 0 K (left panel) and 300 K (right panel). The color squares
denotes the points $T_c$ (red), $T_{c'}$ (blue), $K_{c}$ 
(purple), $K_{v}$ (green). Spectral function are normalized. The color scale bar 
indicates the maximum value in black and the minimum in white.}
\label{spectral-functions}
\end{figure*}

Figure \ref{spectral-functions} shows the spectral function of single-layer
MoS$_2$ for temperature 0 K (left panel) and 300 K (right panel).
An animated represenation of the band
structures for temperature ranging from 0 to 1000 K in step of 100 K can be found in the Supplementary
Informations. Dotted black lines represent
the LDA band structure (without electron-phonon interaction). We have marked
with squares some important points in the band structures, which will be
discussed in more detail below. The bands are no longer a line 
and they acquire a broadening. This broadening is directly related to the
linewidths of each quasiparticle
state. Considering that lifetimes are inversely proportional to linewidths,  
narrow lines are related to long lifetimes, i.e., stable states. On the opposite,
broader states have a stronger interaction with phonons and they have more non-radiative recombination
paths, meaning a shorter lifetime.

We identify very narrow line shapes at the band edges like the valence band states at 
$\bm{K}$ and $\Gamma$ points. In the conduction band we find narrow line shapes at $\bm{K}$ and at the
minimum between $\bm{K}$ and $\Gamma$. Temperature tends to reduce the quasi-particle energy but it does not change significantly
the spectral function, it only moves the maximum to lower energies. The increase of temperature results in a shrinking of the gap. Even at 0 K the gap is 
diminished by 75 meV (with respect to its value calculated without electron-phonon coupling). This is an effect of the zero-point vibrations of the atoms.

The spectral functions of quasi-particle states far from the band edges have a
different aspect. Close to crossings, the bands become blurred, making it
difficult to distinguish individual bands. For instance, the 
conduction band around $\Gamma$ and the crossing close to $\bm{M}$ have a noticeable 
broadening, even at 0 K. The $M$ point shows also broader bands than the $\Gamma$ and the $\bm{K}$ point. The increasing
of temperature blurs even more the reminiscence of the LDA band dispersion. In areas close to $\Gamma$, the band index
becomes almost obsolete and we observe a wide spectral range.\cite{stone_quasiparticle_2006} It is worth to note that 
quasi-particle states are not necessarily broadened peaks centered at the renormalized electron energy. They can also be mixed states which can have a structure very different
from the superposition of the electron and hole density of states. We also expect important 
consequences on the optical properties. Excitons from states in these range of energies (like
the resonant or van-Hove exciton\cite{molina-sanchez_effect_2013,qiu_optical_2013}) should
be affected by the increasing of temperature much more that those coming from band edges.

\begin{figure}
\includegraphics[width=7 cm]{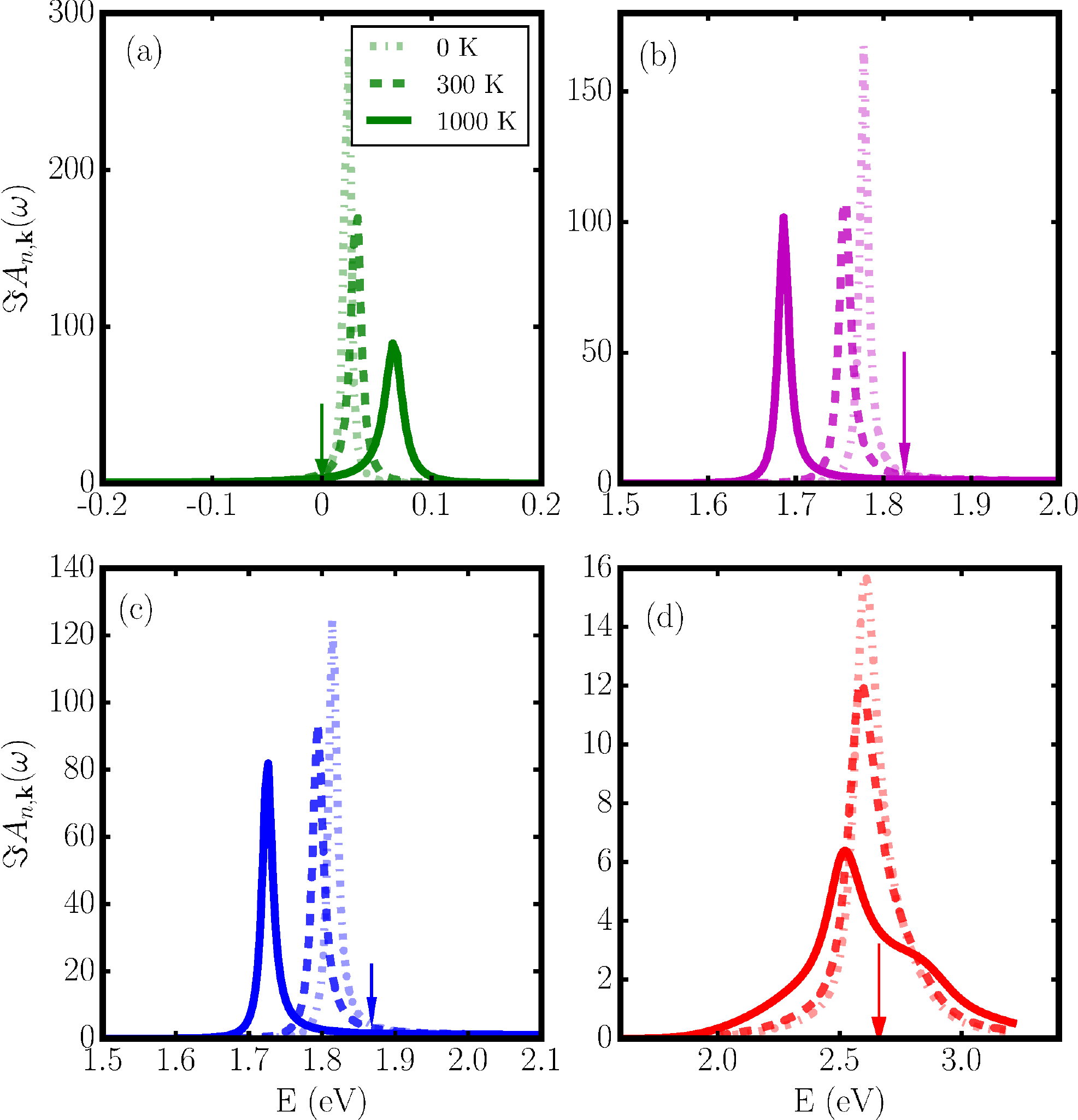}
\caption{Spectral functions for selected band states 
in Fig. \ref{spectral-functions}, $T_c$ (red), $T'_c$ (blue), $K_c$ (magenta) 
and $K_v$ (green), and for $T = 0$ (dotted), 300 (dashed) and 1000 (solid) K. }
\label{sf-selected}
\end{figure}

Figure \ref{sf-selected} represents the spectral function of the quasi-particle
states marked with squares in Fig. \ref{spectral-functions}. We have
selected three temperatures, 0 (dotted), 300 (dashed) and 1000 (solid) K. The arrow indicates the
LDA energy. Even though 1000 K is a very high
temperature for common experiments, it can help us conceptually to discuss the nature of the 
electron-phonon interaction. Panels (a) and (b) of Fig.~\ref{sf-selected} 
show the conduction and valence band extrema at $\bm{K}$. We observe a
shift of the peaks with a slight broadening, but always conserving the Lorentzian shape. 
Regarding the spin-orbit interaction, there is no electron-phonon mediated spin mixing, neither of the conduction nor of the valence
band states at $\bm{K}$.\footnote{By spin mixing we mean that an electron in the VBM
and with momentum $K$ and spin up or mostly up can relax into a state at $K$
with spin down or mostly down.} Our calculation rules out the possibility of
intervalley scattering from the point $K$ to $K'$ at the VBM.  
For 
the conduction band states, the electron-phonon interaction 
conserves the spin degeneracy. 
However, the valence band states are splitted due to the spin-orbit interaction. Comparison with MoS$_2$ ARPES data collected at 80 K is 
a delicate issue.\cite{Miwa2015}  Experimental 
broadening is not exclusively related to 
electron-phonon decay. Nevertheless, the measured spin-orbit splitting of 
145 meV agrees
very well with our calculation of 135 meV.  

Fig.\ref{sf-selected} (c) shows the spectral function of the state in the local minimum between $\bm{K}$
and $\Gamma$. This spectral function has a similar behavior as the cases (a) and (b) but its asymmetry
is stronger. This result is compatible with the exposition of Ref.~\onlinecite{li_intrinsic_2013}, in which transitions
are possible from this local minima to the point $\bm{K}$.

We have found a signature of a potential breakdown of the quasi-particle approximation for some 
states above the band gap. Figure \ref{sf-selected} (d) shows a drastic change 
of the spectral function
due to temperature effects. We have chosen a band close to $\Gamma$, relevant for describing the 
exciton $C$. The shape even at 300 K becomes asymmetric and when we reach 1000 K a secondary
peak emerges. Notice that the high energy peak is separated from the low energy one by an energy far
larger than any phonon in MoS$_2$.\cite{molina-sanchez_phonons_2011} This latter peak appears at higher
energy of the LDA energy, contrary to the others spectral functions. This is a proof of the many-body
character of the new state and of the breakdown of the quasi-particle approximation. The new states cannot be
interpreted any more as an independent sum of electrons and phonon replica. 
The energy separation between the shoulder and the lower peak is bigger than any
phonon frequency. Following Ref. \onlinecite{cannuccia_zero_2012}, the electron
is fragmented in several entangled electron-phonon states, as a result of
virtual transitions not bound to respect the energy conservation. This explains
the appearance of these structures in a wide energy range.\cite{cannuccia_effect_2011} 

We follow the analysis of the temperature dependent electronic structure by investigating
the band gap renormalization. We have seen that at 0 K the band gap of single-layer
MoS$_2$ shrinks. The reason is the uncertainty principle and this is known as the zero-point
motion renormalization (ZPR) effect. At 0 K, atoms cannot be at rest and have zero velocity, there
is a minimum quantum of energy which supplies the vibration which makes possible the 
electron-phonon interaction. Table \ref{gap-table} shows the ZPR for several semiconductors,
calculated in previous works. Single-layer 
MoS$_2$ exhibits a smaller ZPR effect, especially in comparison with Diamond. The wave function
of the conduction and valence band state at $\bm{K}$ are mostly concentrated around the
molybdenum atoms. The large mass of molybdenum reduces the phonon amplitude with the 
consequence of a smaller correction.

\begin{table}
\begin{tabular}{lc}
\hline
\hline
         &  ZPR (meV)    \\
\hline
 MoS$_2$ &   75          \\       
\hline
  Diamond  &  622\cite{antonius_many-body_2014}         \\
  SiC      &  223\cite{monserrat_comparing_2014}           \\
  Si       &  123\cite{monserrat_comparing_2014}         \\
\hline
\hline   
\end{tabular}
\caption{Zero-point motion renormalization of single-layer MoS$_2$, Diamond, SiC and Si.}
\label{gap-table}
\end{table}

In order to shed light on which phonon modes
contribute to the electron-phonon interaction it is useful to calculate 
the Eliashbergh functions.

\begin{multline}
g^2F(\omega)  = \sum_{\lambda \bf q}\left[ \frac{\sum_{n'}|g^{\bf q \lambda}_{nn'\bf k}|N^{-1}_{\bf q}}{\varepsilon_{n \bf k}-\varepsilon_{n' \bf k'}} \right]\delta(\omega-\omega_{\bf q \lambda})
\\ -
 \sum_{\lambda \bf q}\left[2\frac{\sum_{n'}\Lambda^{\bf q \lambda}_{nn'\bf k}N^{-1}_{\bf q}}{\varepsilon_{n \bf k}-\varepsilon_{n' \bf k'}} \right]\delta(\omega-\omega_{\bf q \lambda}).
\end{multline}

\begin{figure*}
\includegraphics[width=14 cm]{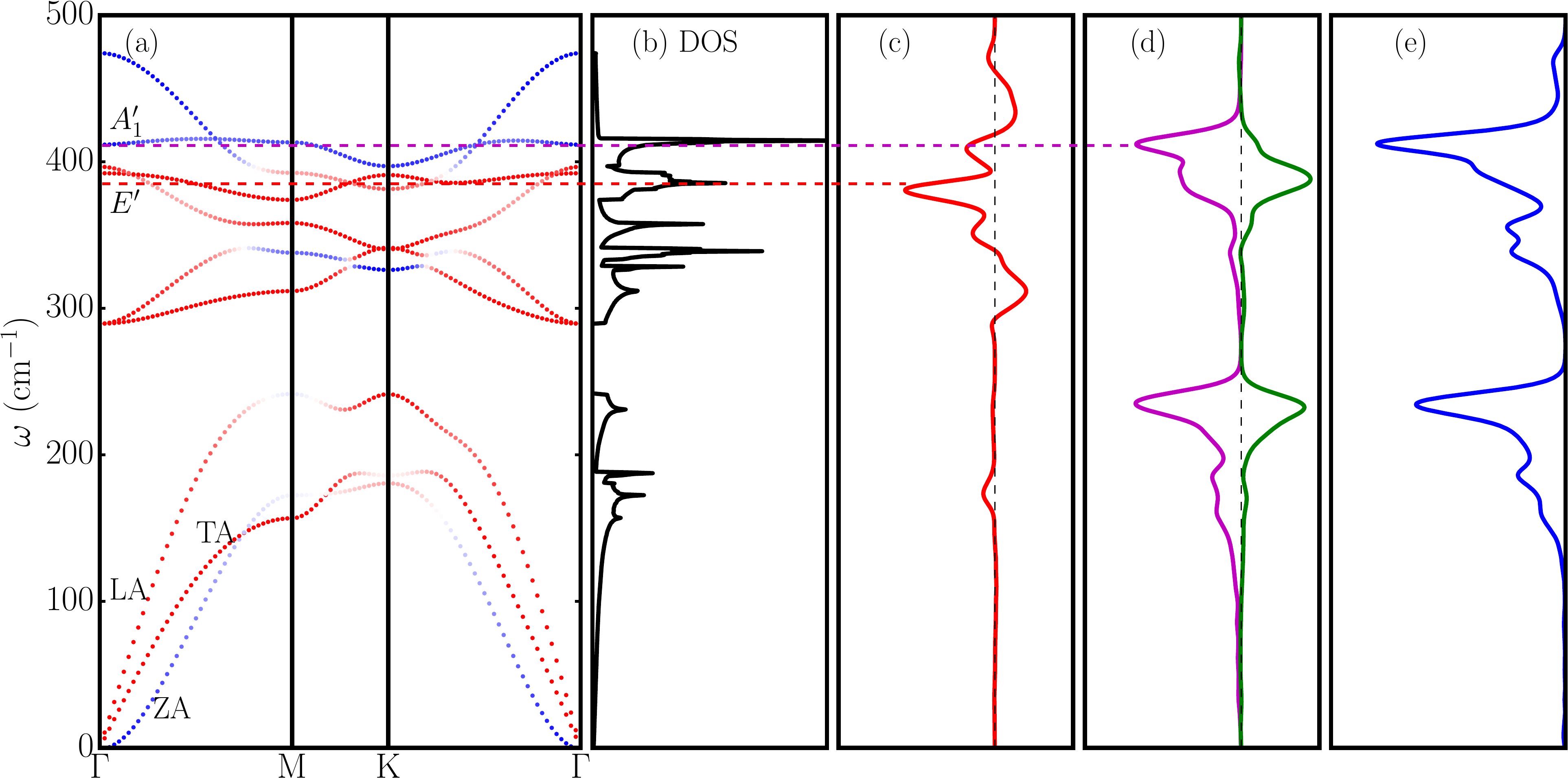}
\caption{Phonon band structure and density of states of single-layer MoS$_2$. Eliasberg functions
of the band states $T_c$, $T'_c$, $K_c$ and $K_v$ (see definitions in the text).}
\label{eliash}
\end{figure*}

Figure \ref{eliash} shows phonon dispersion of single-layer MoS$_2$ (panel a), together with the phonon density of
states (panel b) and the Eliashbergh functions (panels (c),(d),(e))
calculated for the  quasi-particle states  marked with the same color in Figs.
\ref{spectral-functions} and \ref{sf-selected}. In Fig. \ref{eliash} (a) the color 
of the phonon dispersion curves indicate the vibration mode direction: vibrations in-plane are represented by red dots and
out-of-plane vibrations are represented by blue dots. 

A common trend to all the Eliashberg functions is the absence of 
acoustic-phonon contributions close to $\Gamma$. In calculations without the DW term (not
shown here), there is a finite contribution of the Fan Eliasbergh function,
which is removed once the DW is added. Therefore, even though the DW term has 
a small contribution, it is important in order to achieve accurate results. 
The states at $K$ (panel (d))  have
similar Eliashberg functions, almost symmetric. They have opposite sign which results in the shrinking of
the band gap. The main contributions come from
phonons at the edge of the Brillouin zone around 200 cm$^{-1}$ and from optical 
phonons around 400 cm$^{-1}$. In the case of the phonons of the state $T_c$ (e) we find a similar 
Eliashberg function. The high frequency contribution is almost identical
to the one of the state $K_c$.

The Eliashberg function for the state close to $\Gamma$ (panel c)
has a different shape than the others. 
There is no contribution from mid-frequency phonons. The main interaction is due to
optical phonons, around 380 cm$^{-1}$. In this case, the Eliashberg function
changes the sign, crossing several times the zero axis (dotted lines). The Eliasberg functions shown in Fig. \ref{eliash} (d) 
are centered close to the frequency of mode $A_{1g}$, in a frequency range exclusively populated by out-of-plane phonons. In contrast, the function
in panel (c) is built from in-plane phonons close 
to the frequency of the phonon mode $E_{2g}$ 
These findings seem to support the conclusion reached in a recent work  by
Carvalho \textit{et. al}\cite{carvalho_symmetry-dependent_2015} 
on the base of pure symmetry arguments.
Measuring Raman spectra of a MoS$_2$  monolayer,
they concluded that exciton A couples with phonon mode $A_{1g}$ while  exciton C
couples with phonon mode $E_{2g}$  
\cite{molina-sanchez_effect_2013,qiu_optical_2013}.

Since excitons and phonons are particularly complex in TMDs, some deviations can occur as excitons 
are built from many electron and hole states of different electron momentum $\bf k$ and we are just examining a few states. 
We have analyzed the Eliashberg functions at different k points (around
$K$ and  $\Gamma$)
observing a small energy shifts but not drastic changes under small changes of $\bf k$. 
From this result we can affirm that the identification
of Ref. \onlinecite{carvalho_symmetry-dependent_2015} correspond to the excitons A and C. 
A definitive proof
of this statement would consist in calculating the Raman tensor (in resonant conditions)
but this is out of the scope of our work.\cite{gillet_first-principles_2013}

\section{Finite temperature excitonic effects on the optical absorption}

As mentioned in the introduction it is well known that 
temperature not only affects the energies but also the widths of the peaks in the optical spectra of materials. 
Up to now a systematic study of the behaviour of the absorption spectrum of
MoS$_2$ (both as single-layer or bulk) on temperature is still missing.
Only  the low energy A exciton has been measured in photoluminescence
at different temperature.\cite{korn_low-temperature_2011} 
Measurements of the absorption (reflectance) spectra of Mo$S_2$
are usually done at room temperature.

From the theoretical point of view, if electron-phonon interaction
is not taken into account, the {\it ab-initio}
optical spectra are restricted to the use of a
homogeneous \textit{ad-hoc} broadening. Here, 
following Ref. \onlinecite{marini_textitab_2008}, we solve the 
temperature-dependent Bethe-Salpeter equation, where the corresponding 
excitonic Hamiltonian is:  

\begin{multline}
H^{FA}_{ee'hh'} = (E_e+\Delta E_e(T) -E_h- \Delta E_h(T))\delta_{eh,e'h'} \\+ (f_e-f_h)\Xi_{ee'hh'},
\label{exciton-hamil}
\end{multline}

$E_e$ and $E_h$ stand for electron and hole energies, $f_e$ and $f_h$ are the occupations and
$\Xi_{ee'hh'}$ is the Bethe-Salpeter (BS)
kernel.\cite{rohlfing_electron-hole_2000,marini_textitab_2008} The
BS kernel is the sum of the direct and exchange electron-hole
scattering. In a temperature independent formulation, we would calculate
the energies and the BS kernel from DFT with the corresponding GW corrections
to take care of the bandgap underestimation 
inherent to DFT.\cite{Onida2002}
In this work we have used a scissor operator of 0.925 eV and a stretching factor
of 1.2 for the conduction and valence bands. These values were obtained by
comparison of the DFT-LDA band structure with a GW calculation for single-layer
MoS$_2$.\cite{surfscirep_alejandro,shi_quasiparticle_2013} In the temperature-dependent approach we
use the QP eigenvalues obtained from Eq. \ref{qp-energy}, which now are complex
numbers and depend on temperature. As said above, the finite linewidth is given
as the imaginary part of the eigenvalue correction. The 
temperature-dependent BS Eq.
\ref{exciton-hamil} contains a non-hermitian
operator. The excitonic states will thus have a complex energy $E^X(T)$ depending on
temperature and the imaginary part represents the non-radiative
linewidth of the exciton.\cite{marini_textitab_2008} Moreover, temperature
not only changes the energy of the excitonic states, adding the imaginary term
for the linewidth. In systems with a strong electron-lattice interaction,
temperature-dependent excitonic states are a mixture of the excitonic states
from the temperature-independent regime. For instance, in hexagonal boron
nitride, temperature changes dramatically the oscillator strength of the excitons and one
observes temperature-driven transition from dark to bright 
exciton.\cite{marini_textitab_2008} Certainly, the temperature effect will also
depend on the kind of excitons as we will see below. It is also worth to mention that
although at $T=0$ K the BS Eq. does not reduce to the frozen-atom approximation
due to the zero-point vibrations. We only recover the BS equation within the
frozen-atom approximation when the terms $\Delta E_e(T)$ and $\Delta E_h(T)$
are explicitly removed.

The dielectric function\cite{marini_textitab_2008} depends
explicitly on the temperature
\begin{equation}
\varepsilon(\omega,T) \propto \sum_X
|S_X(T)|^2\Im \left({\frac{1}{\omega-E_X(T)}}\right),
\end{equation}
where $S_X(T)$ is the oscillator strength of each exciton. The broadening 
of the excitonic peaks is introduced naturally as the
imaginary part of the exciton energy, without introducing any damping
parameter. It is worth to notice than the linewidth associated with the
electron-electron interaction is negligible in the energy range in which we
study the optical spectra.

\begin{figure}
\includegraphics[width=8 cm]{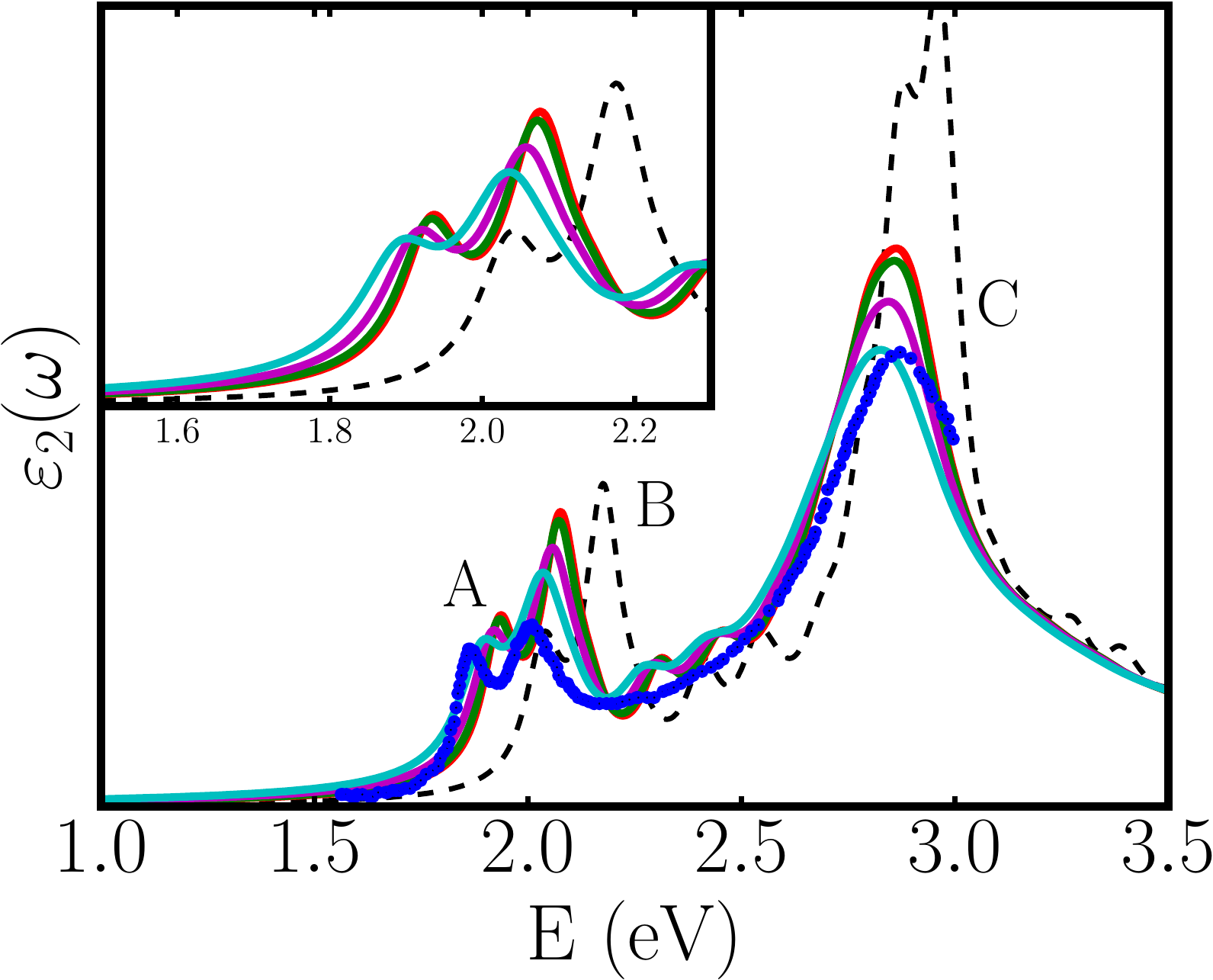}
\caption{Optical spectra of single-layer MoS$_2$ as a function
of temperature. Dashed line represents optical spectra without temperature effects, and the solid line, in increasing intensity temperature of 0, 100, 200 and 300 K (red, green, magenta, cyan). Dots are experimental data 
from Ref. \onlinecite{li_measurement_2014}.}
\label{bse-temp}
\end{figure}

Figure \ref{bse-temp} shows the Bethe-Salpeter spectra calculated without electron-phonon interaction (black dashed line), 
at 0, 100, 200 and 300 K (red, green, magenta, and cyan respectively), and with
dots the experimental data at room temperature from Ref.
\onlinecite{li_measurement_2014}. The change on the electronic states due to
temperature has a repercussion on the excitons and 
on the optical spectra. We have calculated the Bethe-Salpeter spectra in a $30\times 30\times 1$ $\bf{k}-$grid, for 4 conduction band states and
2 valence band states. The rest of convergence criteria can be found 
elsewhere.\cite{molina-sanchez_effect_2013} The temperature correction to the quasi-particle states have been done
in the same $\bf{k}$-grid and following the previous convergence criteria with respect to the number of bands
and $\bf q$ points.

First, the A and B excitons are shifted down in energy but the intensity is rather constant. The A peak
is slightly narrower than the B peak, in agreement with the experiments. The B excitons is build
mainly from the second valence band maximum, which has more non-radiative paths
for recombination than the A exciton. The behavior of the C exciton is
drastically different to that of A and B excitons. The C exciton comes from
transitions close to the $\Gamma$. In this region of the band structure
the electron-phonon interaction alters substantially the
electronic states but not the energy. The intensity drops remarkably from
the BS spectrum in absence of electron-phonon interaction. It is worth to notice than we have used an
homogeneous broadening of 50 meV for the BS spectra without electron-phonon
interaction. The increasing
of temperature reduces C exciton intensity with a faster pace than
in the case of the others excitons and also increases the width.
Another effect is the collapse of the multi-peak structure at
the LDA spectra in one broad peak. The result is consistent with
the spectral functions of Figs. \ref{spectral-functions}
and \ref{sf-selected}.

\begin{figure}
\includegraphics[width=8 cm]{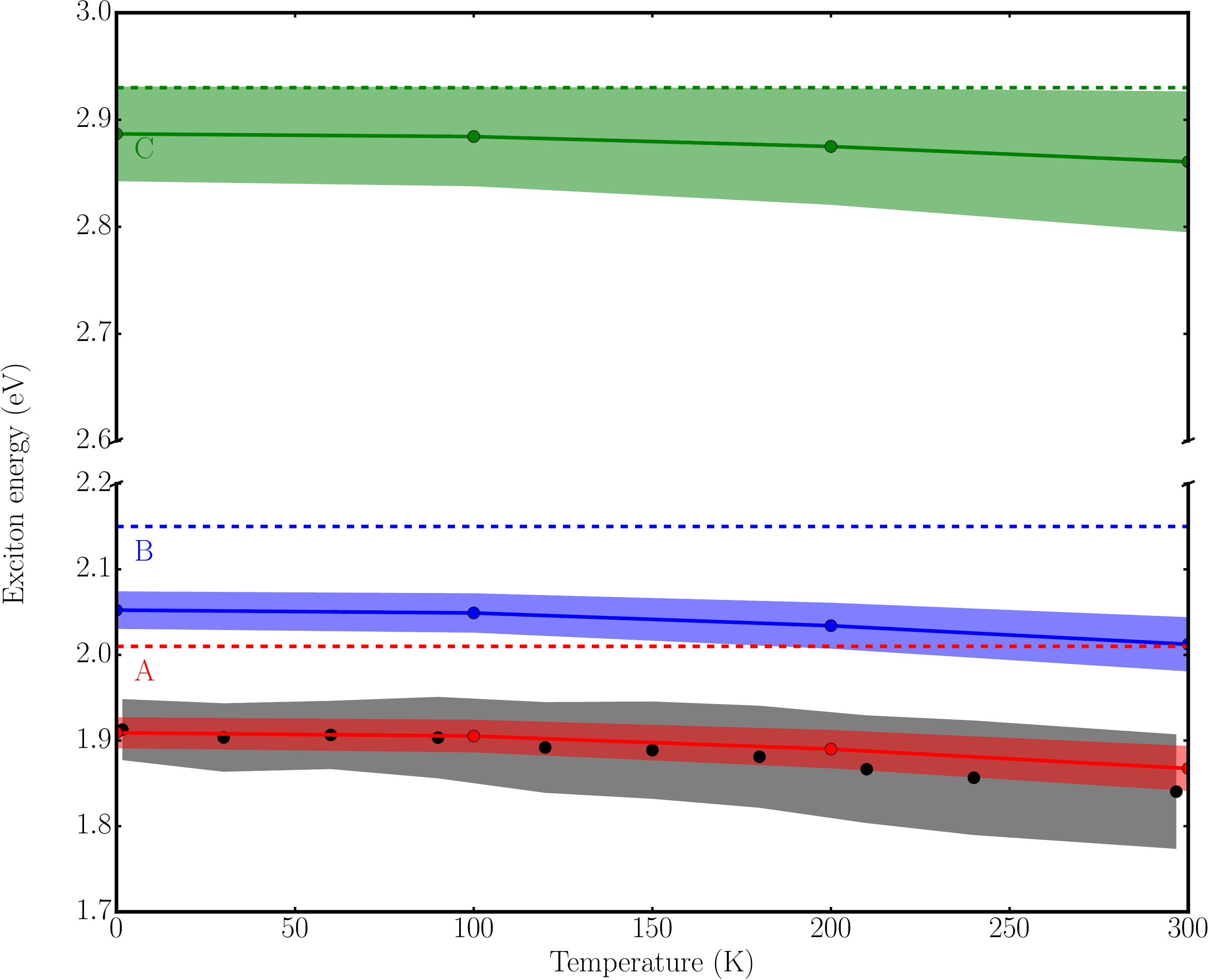}
\caption{Exciton energy as a function of temperature (solid lines). Shadow region shows the width of each excitons. Dashed line
marks the exciton energy without electron-phonon interaction. The
photoluminescence data of Ref. \onlinecite{korn_low-temperature_2011} have been
represented by black dots and the FWHM with 
the gray area.}
\label{binding-temp}
\end{figure}

In order to see in a clearer way the temperature effects on the exciton energies, 
Figure \ref{binding-temp} shows the exciton energy as function of temperature. The dashed area stands for the width of
every excitonic state. The dashed line represents the exciton energy without electron-phonon interaction. All the
excitonic states decreases their energy with increasing temperature but not with the same pace. The biggest correction to
the energy of the excitons A and B is made by the ZPR, being very similar to
both states (75 meV). We can see that the width of
these states is almost constant with the increasing of temperature, being only slightly bigger for the B exciton (44 meV vs. 36 meV). In
the case of the C excitons we have a surprising behaviour. We expected a larger ZPR correction, proportional to the width. While the width is already large, 88 meV) 
at 0 K and 132 meV at 300 K, the temperature increasing does not imply a strong correction of the 
excitonic energy, which remains rather constant. From the spectral functions of Figs. \ref{spectral-functions} and
\ref{sf-selected}(d) we can see that the states close to $\Gamma$ exhibit a remarkably increasing of the broadening but it seems that
more or less centred at the same energy. We have added the
photoluminescence results of Ref. \onlinecite{korn_low-temperature_2011},
representing the full width at half-maximum (FWHM) with the gray dashed area. The 
energies show a good agreement at low temperatures and diverge slightly starting at 
200 K. The main causes of this disagreement is the thermal 
expansion, not included in our calculations. Regarding the widths, both
experimental and theoretical values increase with temperature but the
experimental to a larger extend. This suggests the contribution of more
processes like the radiative recombination, has larger lifetimes than
the carrier-phonon scattering processes described here. From the theoretical
results we can infer that the band-gap dependence on temperature is dictated mainly
by EP interaction, whereas linewidths are influenced by other processes like
radiative recombination or defects scattering.\cite{korn_low-temperature_2011}

The comparison with the experimental data is rather satisfactory. We
can explain the broadening of the C peak as the coupling of electron
with lattice vibrations. On the another side, the A and B peaks
compare also well. The approximation made for modeling the C
exciton seems to be valid, at least to give a qualitative explanation
of the spectral width. It is worth to note that modelling temperature effects on
the optical properties
cannot rely only on the thermal 
expansion,\cite{Tongay2012} especially at temperatures below 200 K, where
the thermal expansion is small. Moreover, only by taking into account the
electron-phonon interaction we can calculate non-radiative linewidths and
interpret some data from photoluminescence spectra such as the FWHM 
or the broadening of the
optical spectra.

\section{Conclusions}

We have presented, to our knowledge, the first calculations of temperature
effects on the electronic structure and the optical properties of a 2D 
material in the presence of spin-orbit coupling. 
For this purpose, we have calculated the electron-phonon matrix
elements and the temperature-dependent spectral function using                 
full spinorial wave functions. We have used single-layer MoS$_2$ as a test 
material and we expect that this work is a good basis for studies in other
monolayer TMDs and in multi-layer MoS$_2$. The electron-phonon interaction
serves also to understand the behaviour of the resonant Raman
spectroscopy. The Eliashberg functions evaluate the exciton-phonon coupling and we can identify which 
excitons will couple to each phonon mode.
We have also discovered a different behaviour with temperature for the two
kind of excitons existing in MoS2. First, excitons from the band edges (bound excitons A and B)
are down-shifted in energy when temperature increases and the small linewidth does not
change significantly. In the case of resonant excitons (C exciton) the situation is more complex. Our calculations show
that bands far from the bandgap have more non-radiative paths available for decaying, as
the electron states occupy a wider energy range. Consequently, the non-radiative
linewidth is strongly
affected by the increase of temperature. The overall result is an optical absorption with a characteristic
inhomogenous broadening, with a C peak much broader than A and B peaks. Our theoretical spectra agrees well
with recent experimental measurements. With this contribution we show the importance of temperature effects, determined by electron-phonon
coupling, for a more realistic approach to the optical properties of semiconductor 2D materials.

\section{Acknowledgements}
A. M.-S.\ and L.W. acknowledge support by the National Research Fund, 
Luxembourg (Projects C14/MS/773152/FAST-2DMAT and  INTER/ANR/13/20/NANOTMD).
M. Palummo acknowledges the support received from
the European Science Foundation (ESF) for the activity entitled 'Advanced
Concepts in {\it Ab-initio} Simulations of Materials' and EC for the RISE Project
CoExAN GA644076. We acknowledge helpful discussions with M. Calandra about 
the calculation of electron-phonon matrix elements with spin-orbit coupling.
AM acknowledges financial support by the Futuro in Ricerca grant No.
RBFR12SW0J of the Italian Ministry of Education, University and Research MIUR, the European Union
project MaX Materials design at the eXascale H2020-EINFRA-2015-1, Grant agreement
n. 676598 and Nanoscience Foundries and Fine Analysis - Europe H2020-INFRAIA-2014-2015,
Grant agreement n. 654360.

\bibliographystyle{apsrev4-1}

\begin{thebibliography}{47}%
\makeatletter
\providecommand \@ifxundefined [1]{%
 \@ifx{#1\undefined}
}%
\providecommand \@ifnum [1]{%
 \ifnum #1\expandafter \@firstoftwo
 \else \expandafter \@secondoftwo
 \fi
}%
\providecommand \@ifx [1]{%
 \ifx #1\expandafter \@firstoftwo
 \else \expandafter \@secondoftwo
 \fi
}%
\providecommand \natexlab [1]{#1}%
\providecommand \enquote  [1]{``#1''}%
\providecommand \bibnamefont  [1]{#1}%
\providecommand \bibfnamefont [1]{#1}%
\providecommand \citenamefont [1]{#1}%
\providecommand \href@noop [0]{\@secondoftwo}%
\providecommand \href [0]{\begingroup \@sanitize@url \@href}%
\providecommand \@href[1]{\@@startlink{#1}\@@href}%
\providecommand \@@href[1]{\endgroup#1\@@endlink}%
\providecommand \@sanitize@url [0]{\catcode `\\12\catcode `\$12\catcode
  `\&12\catcode `\#12\catcode `\^12\catcode `\_12\catcode `\%12\relax}%
\providecommand \@@startlink[1]{}%
\providecommand \@@endlink[0]{}%
\providecommand \url  [0]{\begingroup\@sanitize@url \@url }%
\providecommand \@url [1]{\endgroup\@href {#1}{\urlprefix }}%
\providecommand \urlprefix  [0]{URL }%
\providecommand \Eprint [0]{\href }%
\providecommand \doibase [0]{http://dx.doi.org/}%
\providecommand \selectlanguage [0]{\@gobble}%
\providecommand \bibinfo  [0]{\@secondoftwo}%
\providecommand \bibfield  [0]{\@secondoftwo}%
\providecommand \translation [1]{[#1]}%
\providecommand \BibitemOpen [0]{}%
\providecommand \bibitemStop [0]{}%
\providecommand \bibitemNoStop [0]{.\EOS\space}%
\providecommand \EOS [0]{\spacefactor3000\relax}%
\providecommand \BibitemShut  [1]{\csname bibitem#1\endcsname}%
\let\auto@bib@innerbib\@empty
\bibitem [{\citenamefont {Geim}\ and\ \citenamefont
  {Novoselov}(2007)}]{geim_rise_2007}%
  \BibitemOpen
  \bibfield  {author} {\bibinfo {author} {\bibfnamefont {A.~K.}\ \bibnamefont
  {Geim}}\ and\ \bibinfo {author} {\bibfnamefont {K.~S.}\ \bibnamefont
  {Novoselov}},\ }\href {\doibase 10.1038/nmat1849} {\bibfield  {journal}
  {\bibinfo  {journal} {Nature Materials}\ }\textbf {\bibinfo {volume} {6}},\
  \bibinfo {pages} {183} (\bibinfo {year} {2007})}\BibitemShut {NoStop}%
\bibitem [{\citenamefont {Roy}\ \emph {et~al.}(2013)\citenamefont {Roy},
  \citenamefont {Padmanabhan}, \citenamefont {Goswami}, \citenamefont {Sai},
  \citenamefont {Ramalingam}, \citenamefont {Raghavan},\ and\ \citenamefont
  {Ghosh}}]{roy_graphene-mos2_2013}%
  \BibitemOpen
  \bibfield  {author} {\bibinfo {author} {\bibfnamefont {K.}~\bibnamefont
  {Roy}}, \bibinfo {author} {\bibfnamefont {M.}~\bibnamefont {Padmanabhan}},
  \bibinfo {author} {\bibfnamefont {S.}~\bibnamefont {Goswami}}, \bibinfo
  {author} {\bibfnamefont {T.~P.}\ \bibnamefont {Sai}}, \bibinfo {author}
  {\bibfnamefont {G.}~\bibnamefont {Ramalingam}}, \bibinfo {author}
  {\bibfnamefont {S.}~\bibnamefont {Raghavan}}, \ and\ \bibinfo {author}
  {\bibfnamefont {A.}~\bibnamefont {Ghosh}},\ }\href {\doibase
  10.1038/nnano.2013.206} {\bibfield  {journal} {\bibinfo  {journal} {Nature
  Nanotechnology}\ }\textbf {\bibinfo {volume} {8}},\ \bibinfo {pages} {826}
  (\bibinfo {year} {2013})}\BibitemShut {NoStop}%
\bibitem [{\citenamefont {Lopez-Sanchez}\ \emph {et~al.}(2014)\citenamefont
  {Lopez-Sanchez}, \citenamefont {Alarcon~Llado}, \citenamefont {Koman},
  \citenamefont {Fontcuberta~i Morral}, \citenamefont {Radenovic},\ and\
  \citenamefont {Kis}}]{lopez-sanchez_light_2014}%
  \BibitemOpen
  \bibfield  {author} {\bibinfo {author} {\bibfnamefont {O.}~\bibnamefont
  {Lopez-Sanchez}}, \bibinfo {author} {\bibfnamefont {E.}~\bibnamefont
  {Alarcon~Llado}}, \bibinfo {author} {\bibfnamefont {V.}~\bibnamefont
  {Koman}}, \bibinfo {author} {\bibfnamefont {A.}~\bibnamefont {Fontcuberta~i
  Morral}}, \bibinfo {author} {\bibfnamefont {A.}~\bibnamefont {Radenovic}}, \
  and\ \bibinfo {author} {\bibfnamefont {A.}~\bibnamefont {Kis}},\ }\href
  {\doibase 10.1021/nn500480u} {\bibfield  {journal} {\bibinfo  {journal} {ACS
  Nano}\ }\textbf {\bibinfo {volume} {8}},\ \bibinfo {pages} {3042} (\bibinfo
  {year} {2014})}\BibitemShut {NoStop}%
\bibitem [{\citenamefont {Radisavljevic}\ \emph {et~al.}(2011)\citenamefont
  {Radisavljevic}, \citenamefont {Radenovic}, \citenamefont {Brivio},
  \citenamefont {Giacometti},\ and\ \citenamefont
  {Kis}}]{radisavljevic_single-layer_2011}%
  \BibitemOpen
  \bibfield  {author} {\bibinfo {author} {\bibfnamefont {B.}~\bibnamefont
  {Radisavljevic}}, \bibinfo {author} {\bibfnamefont {A.}~\bibnamefont
  {Radenovic}}, \bibinfo {author} {\bibfnamefont {J.}~\bibnamefont {Brivio}},
  \bibinfo {author} {\bibfnamefont {V.}~\bibnamefont {Giacometti}}, \ and\
  \bibinfo {author} {\bibfnamefont {A.}~\bibnamefont {Kis}},\ }\href {\doibase
  10.1038/nnano.2010.279} {\bibfield  {journal} {\bibinfo  {journal} {Nature
  Nanotechnology}\ }\textbf {\bibinfo {volume} {6}},\ \bibinfo {pages} {147}
  (\bibinfo {year} {2011})}\BibitemShut {NoStop}%
\bibitem [{\citenamefont {Mak}\ \emph {et~al.}(2014)\citenamefont {Mak},
  \citenamefont {McGill}, \citenamefont {Park},\ and\ \citenamefont
  {McEuen}}]{mak_valley_2014}%
  \BibitemOpen
  \bibfield  {author} {\bibinfo {author} {\bibfnamefont {K.~F.}\ \bibnamefont
  {Mak}}, \bibinfo {author} {\bibfnamefont {K.~L.}\ \bibnamefont {McGill}},
  \bibinfo {author} {\bibfnamefont {J.}~\bibnamefont {Park}}, \ and\ \bibinfo
  {author} {\bibfnamefont {P.~L.}\ \bibnamefont {McEuen}},\ }\href {\doibase
  10.1126/science.1250140} {\bibfield  {journal} {\bibinfo  {journal}
  {Science}\ }\textbf {\bibinfo {volume} {344}},\ \bibinfo {pages} {1489}
  (\bibinfo {year} {2014})}\BibitemShut {NoStop}%
\bibitem [{\citenamefont {Hong}\ \emph {et~al.}(2014)\citenamefont {Hong},
  \citenamefont {Kim}, \citenamefont {Shi}, \citenamefont {Zhang},
  \citenamefont {Jin}, \citenamefont {Sun}, \citenamefont {Tongay},
  \citenamefont {Wu}, \citenamefont {Zhang},\ and\ \citenamefont
  {Wang}}]{hong_ultrafast_2014}%
  \BibitemOpen
  \bibfield  {author} {\bibinfo {author} {\bibfnamefont {X.}~\bibnamefont
  {Hong}}, \bibinfo {author} {\bibfnamefont {J.}~\bibnamefont {Kim}}, \bibinfo
  {author} {\bibfnamefont {S.-F.}\ \bibnamefont {Shi}}, \bibinfo {author}
  {\bibfnamefont {Y.}~\bibnamefont {Zhang}}, \bibinfo {author} {\bibfnamefont
  {C.}~\bibnamefont {Jin}}, \bibinfo {author} {\bibfnamefont {Y.}~\bibnamefont
  {Sun}}, \bibinfo {author} {\bibfnamefont {S.}~\bibnamefont {Tongay}},
  \bibinfo {author} {\bibfnamefont {J.}~\bibnamefont {Wu}}, \bibinfo {author}
  {\bibfnamefont {Y.}~\bibnamefont {Zhang}}, \ and\ \bibinfo {author}
  {\bibfnamefont {F.}~\bibnamefont {Wang}},\ }\href {\doibase
  10.1038/nnano.2014.167} {\bibfield  {journal} {\bibinfo  {journal} {Nature
  Nanotechnology}\ }\textbf {\bibinfo {volume} {9}},\ \bibinfo {pages} {682}
  (\bibinfo {year} {2014})}\BibitemShut {NoStop}%
\bibitem [{\citenamefont {Yu}\ \emph {et~al.}(2015)\citenamefont {Yu},
  \citenamefont {Cui}, \citenamefont {Xu},\ and\ \citenamefont
  {Yao}}]{yu_valley_2015}%
  \BibitemOpen
  \bibfield  {author} {\bibinfo {author} {\bibfnamefont {H.}~\bibnamefont
  {Yu}}, \bibinfo {author} {\bibfnamefont {X.}~\bibnamefont {Cui}}, \bibinfo
  {author} {\bibfnamefont {X.}~\bibnamefont {Xu}}, \ and\ \bibinfo {author}
  {\bibfnamefont {W.}~\bibnamefont {Yao}},\ }\href {\doibase
  10.1093/nsr/nwu078} {\bibfield  {journal} {\bibinfo  {journal} {National
  Science Review}\ }\textbf {\bibinfo {volume} {2}},\ \bibinfo {pages} {57}
  (\bibinfo {year} {2015})}\BibitemShut {NoStop}%
\bibitem [{\citenamefont {Lembke}\ and\ \citenamefont
  {Kis}(2012)}]{lembke_breakdown_2012}%
  \BibitemOpen
  \bibfield  {author} {\bibinfo {author} {\bibfnamefont {D.}~\bibnamefont
  {Lembke}}\ and\ \bibinfo {author} {\bibfnamefont {A.}~\bibnamefont {Kis}},\
  }\href {\doibase 10.1021/nn303772b} {\bibfield  {journal} {\bibinfo
  {journal} {ACS Nano}\ }\textbf {\bibinfo {volume} {6}},\ \bibinfo {pages}
  {10070} (\bibinfo {year} {2012})}\BibitemShut {NoStop}%
\bibitem [{\citenamefont {Komsa}\ and\ \citenamefont
  {Krasheninnikov}(2012)}]{Komsa2012}%
  \BibitemOpen
  \bibfield  {author} {\bibinfo {author} {\bibfnamefont {H.-P.}\ \bibnamefont
  {Komsa}}\ and\ \bibinfo {author} {\bibfnamefont {A.~V.}\ \bibnamefont
  {Krasheninnikov}},\ }\href {\doibase 10.1103/PhysRevB.86.241201} {\bibfield
  {journal} {\bibinfo  {journal} {Phys. Rev. B}\ }\textbf {\bibinfo {volume}
  {86}},\ \bibinfo {pages} {241201} (\bibinfo {year} {2012})}\BibitemShut
  {NoStop}%
\bibitem [{\citenamefont {Molina-S\'anchez}\ \emph {et~al.}(2013)\citenamefont
  {Molina-S\'anchez}, \citenamefont {Sangalli}, \citenamefont {Hummer},
  \citenamefont {Marini},\ and\ \citenamefont
  {Wirtz}}]{molina-sanchez_effect_2013}%
  \BibitemOpen
  \bibfield  {author} {\bibinfo {author} {\bibfnamefont {A.}~\bibnamefont
  {Molina-S\'anchez}}, \bibinfo {author} {\bibfnamefont {D.}~\bibnamefont
  {Sangalli}}, \bibinfo {author} {\bibfnamefont {K.}~\bibnamefont {Hummer}},
  \bibinfo {author} {\bibfnamefont {A.}~\bibnamefont {Marini}}, \ and\ \bibinfo
  {author} {\bibfnamefont {L.}~\bibnamefont {Wirtz}},\ }\href {\doibase
  10.1103/PhysRevB.88.045412} {\bibfield  {journal} {\bibinfo  {journal}
  {Physical Review B}\ }\textbf {\bibinfo {volume} {88}},\ \bibinfo {pages}
  {045412} (\bibinfo {year} {2013})}\BibitemShut {NoStop}%
\bibitem [{\citenamefont {Qiu}\ \emph {et~al.}(2013)\citenamefont {Qiu},
  \citenamefont {da~Jornada},\ and\ \citenamefont {Louie}}]{qiu_optical_2013}%
  \BibitemOpen
  \bibfield  {author} {\bibinfo {author} {\bibfnamefont {D.~Y.}\ \bibnamefont
  {Qiu}}, \bibinfo {author} {\bibfnamefont {F.~H.}\ \bibnamefont {da~Jornada}},
  \ and\ \bibinfo {author} {\bibfnamefont {S.~G.}\ \bibnamefont {Louie}},\
  }\href {\doibase 10.1103/PhysRevLett.111.216805} {\bibfield  {journal}
  {\bibinfo  {journal} {Physical Review Letters}\ }\textbf {\bibinfo {volume}
  {111}},\ \bibinfo {pages} {216805} (\bibinfo {year} {2013})}\BibitemShut
  {NoStop}%
\bibitem [{\citenamefont {Komsa}\ and\ \citenamefont
  {Krasheninnikov}(2013)}]{Komsa2013}%
  \BibitemOpen
  \bibfield  {author} {\bibinfo {author} {\bibfnamefont {H.-P.}\ \bibnamefont
  {Komsa}}\ and\ \bibinfo {author} {\bibfnamefont {A.~V.}\ \bibnamefont
  {Krasheninnikov}},\ }\href {\doibase 10.1103/PhysRevB.88.085318} {\bibfield
  {journal} {\bibinfo  {journal} {Phys. Rev. B}\ }\textbf {\bibinfo {volume}
  {88}},\ \bibinfo {pages} {085318} (\bibinfo {year} {2013})}\BibitemShut
  {NoStop}%
\bibitem [{\citenamefont {Molina-S\'{a}nchez}\ \emph
  {et~al.}(2015)\citenamefont {Molina-S\'{a}nchez}, \citenamefont {Hummer},\
  and\ \citenamefont {Wirtz}}]{surfscirep_alejandro}%
  \BibitemOpen
  \bibfield  {author} {\bibinfo {author} {\bibfnamefont {A.}~\bibnamefont
  {Molina-S\'{a}nchez}}, \bibinfo {author} {\bibfnamefont {K.}~\bibnamefont
  {Hummer}}, \ and\ \bibinfo {author} {\bibfnamefont {L.}~\bibnamefont
  {Wirtz}},\ }\href {\doibase http://dx.doi.org/10.1016/j.surfrep.2015.10.001}
  {\bibfield  {journal} {\bibinfo  {journal} {Surface Science Reports}\
  }\textbf {\bibinfo {volume} {70}},\ \bibinfo {pages} {554 } (\bibinfo {year}
  {2015})}\BibitemShut {NoStop}%
\bibitem [{\citenamefont {Wolpert}\ and\ \citenamefont
  {Ampadu}(2012)}]{wolpert_temperature_2012}%
  \BibitemOpen
  \bibfield  {author} {\bibinfo {author} {\bibfnamefont {D.}~\bibnamefont
  {Wolpert}}\ and\ \bibinfo {author} {\bibfnamefont {P.}~\bibnamefont
  {Ampadu}}\ }(\bibinfo  {publisher} {Springer New York},\ \bibinfo {year}
  {2012})\ pp.\ \bibinfo {pages} {15--33}\BibitemShut {NoStop}%
\bibitem [{\citenamefont {Lautenschlager}\ \emph {et~al.}(1985)\citenamefont
  {Lautenschlager}, \citenamefont {Allen},\ and\ \citenamefont
  {Cardona}}]{lautenschlager_temperature_1985}%
  \BibitemOpen
  \bibfield  {author} {\bibinfo {author} {\bibfnamefont {P.}~\bibnamefont
  {Lautenschlager}}, \bibinfo {author} {\bibfnamefont {P.~B.}\ \bibnamefont
  {Allen}}, \ and\ \bibinfo {author} {\bibfnamefont {M.}~\bibnamefont
  {Cardona}},\ }\href {\doibase 10.1103/PhysRevB.31.2163} {\bibfield  {journal}
  {\bibinfo  {journal} {Physical Review B}\ }\textbf {\bibinfo {volume} {31}},\
  \bibinfo {pages} {2163} (\bibinfo {year} {1985})}\BibitemShut {NoStop}%
\bibitem [{\citenamefont {Lautenschlager}\ \emph {et~al.}(1987)\citenamefont
  {Lautenschlager}, \citenamefont {Garriga}, \citenamefont {Vina},\ and\
  \citenamefont {Cardona}}]{lautenschlager_temperature_1987}%
  \BibitemOpen
  \bibfield  {author} {\bibinfo {author} {\bibfnamefont {P.}~\bibnamefont
  {Lautenschlager}}, \bibinfo {author} {\bibfnamefont {M.}~\bibnamefont
  {Garriga}}, \bibinfo {author} {\bibfnamefont {L.}~\bibnamefont {Vina}}, \
  and\ \bibinfo {author} {\bibfnamefont {M.}~\bibnamefont {Cardona}},\ }\href
  {\doibase 10.1103/PhysRevB.36.4821} {\bibfield  {journal} {\bibinfo
  {journal} {Physical Review B}\ }\textbf {\bibinfo {volume} {36}},\ \bibinfo
  {pages} {4821} (\bibinfo {year} {1987})}\BibitemShut {NoStop}%
\bibitem [{\citenamefont {Miwa}\ \emph {et~al.}(2015)\citenamefont {Miwa},
  \citenamefont {Sorensen}, \citenamefont {Dendzik}, \citenamefont {Cabo},
  \citenamefont {Bianchi}, \citenamefont {Lauritsen},\ and\ \citenamefont
  {Hofmann}}]{Miwa2015}%
  \BibitemOpen
  \bibfield  {author} {\bibinfo {author} {\bibfnamefont {S.}~\bibnamefont
  {Miwa}, \bibfnamefont {J.A.~andUlstrup}}, \bibinfo {author} {\bibfnamefont
  {S.}~\bibnamefont {Sorensen}}, \bibinfo {author} {\bibfnamefont
  {M.}~\bibnamefont {Dendzik}}, \bibinfo {author} {\bibfnamefont
  {A.}~\bibnamefont {Cabo}}, \bibinfo {author} {\bibfnamefont {M.}~\bibnamefont
  {Bianchi}}, \bibinfo {author} {\bibfnamefont {J.}~\bibnamefont {Lauritsen}},
  \ and\ \bibinfo {author} {\bibfnamefont {P.}~\bibnamefont {Hofmann}},\ }\href
  {\doibase 10.1103/PhysRevLett.114.046802} {\bibfield  {journal} {\bibinfo
  {journal} {Phys. Rev. Lett.}\ }\textbf {\bibinfo {volume} {114}},\ \bibinfo
  {pages} {046802} (\bibinfo {year} {2015})}\BibitemShut {NoStop}%
\bibitem [{\citenamefont {Cardona}(2005)}]{cardona_electronphonon_2005}%
  \BibitemOpen
  \bibfield  {author} {\bibinfo {author} {\bibfnamefont {M.}~\bibnamefont
  {Cardona}},\ }\href {\doibase 10.1016/j.ssc.2004.10.028} {\bibfield
  {journal} {\bibinfo  {journal} {Solid State Communications}\ }\textbf
  {\bibinfo {volume} {133}},\ \bibinfo {pages} {3} (\bibinfo {year}
  {2005})}\BibitemShut {NoStop}%
\bibitem [{\citenamefont {Giustino}\ \emph {et~al.}(2010)\citenamefont
  {Giustino}, \citenamefont {Louie},\ and\ \citenamefont
  {Cohen}}]{giustino_electron-phonon_2010}%
  \BibitemOpen
  \bibfield  {author} {\bibinfo {author} {\bibfnamefont {F.}~\bibnamefont
  {Giustino}}, \bibinfo {author} {\bibfnamefont {S.~G.}\ \bibnamefont {Louie}},
  \ and\ \bibinfo {author} {\bibfnamefont {M.~L.}\ \bibnamefont {Cohen}},\
  }\href {\doibase 10.1103/PhysRevLett.105.265501} {\bibfield  {journal}
  {\bibinfo  {journal} {Physical Review Letters}\ }\textbf {\bibinfo {volume}
  {105}},\ \bibinfo {pages} {265501} (\bibinfo {year} {2010})}\BibitemShut
  {NoStop}%
\bibitem [{\citenamefont {Moser}\ \emph {et~al.}(2013)\citenamefont {Moser},
  \citenamefont {Moreschini}, \citenamefont {Jaimovi}, \citenamefont {Barisic},
  \citenamefont {Berger}, \citenamefont {Magrez}, \citenamefont {Chang},
  \citenamefont {Kim}, \citenamefont {Bostwick}, \citenamefont {Rotenberg},
  \citenamefont {Forr},\ and\ \citenamefont {Grioni}}]{moser_tunable_2013}%
  \BibitemOpen
  \bibfield  {author} {\bibinfo {author} {\bibfnamefont {S.}~\bibnamefont
  {Moser}}, \bibinfo {author} {\bibfnamefont {L.}~\bibnamefont {Moreschini}},
  \bibinfo {author} {\bibfnamefont {J.}~\bibnamefont {Jaimovi}}, \bibinfo
  {author} {\bibfnamefont {O.~S.}\ \bibnamefont {Barisic}}, \bibinfo {author}
  {\bibfnamefont {H.}~\bibnamefont {Berger}}, \bibinfo {author} {\bibfnamefont
  {A.}~\bibnamefont {Magrez}}, \bibinfo {author} {\bibfnamefont {Y.~J.}\
  \bibnamefont {Chang}}, \bibinfo {author} {\bibfnamefont {K.~S.}\ \bibnamefont
  {Kim}}, \bibinfo {author} {\bibfnamefont {A.}~\bibnamefont {Bostwick}},
  \bibinfo {author} {\bibfnamefont {E.}~\bibnamefont {Rotenberg}}, \bibinfo
  {author} {\bibfnamefont {L.}~\bibnamefont {Forr}}, \ and\ \bibinfo {author}
  {\bibfnamefont {M.}~\bibnamefont {Grioni}},\ }\href {\doibase
  10.1103/PhysRevLett.110.196403} {\bibfield  {journal} {\bibinfo  {journal}
  {Physical Review Letters}\ }\textbf {\bibinfo {volume} {110}},\ \bibinfo
  {pages} {196403} (\bibinfo {year} {2013})}\BibitemShut {NoStop}%
\bibitem [{\citenamefont {Kaasbjerg}\ \emph {et~al.}(2012)\citenamefont
  {Kaasbjerg}, \citenamefont {Thygesen},\ and\ \citenamefont
  {Jacobsen}}]{kaasbjerg_phonon-limited_2012}%
  \BibitemOpen
  \bibfield  {author} {\bibinfo {author} {\bibfnamefont {K.}~\bibnamefont
  {Kaasbjerg}}, \bibinfo {author} {\bibfnamefont {K.~S.}\ \bibnamefont
  {Thygesen}}, \ and\ \bibinfo {author} {\bibfnamefont {K.~W.}\ \bibnamefont
  {Jacobsen}},\ }\href {\doibase 10.1103/PhysRevB.85.115317} {\bibfield
  {journal} {\bibinfo  {journal} {Physical Review B}\ }\textbf {\bibinfo
  {volume} {85}},\ \bibinfo {pages} {115317} (\bibinfo {year}
  {2012})}\BibitemShut {NoStop}%
\bibitem [{\citenamefont {Li}\ \emph {et~al.}(2013{\natexlab{a}})\citenamefont
  {Li}, \citenamefont {Carrete},\ and\ \citenamefont
  {Mingo}}]{li_thermal_2013}%
  \BibitemOpen
  \bibfield  {author} {\bibinfo {author} {\bibfnamefont {W.}~\bibnamefont
  {Li}}, \bibinfo {author} {\bibfnamefont {J.}~\bibnamefont {Carrete}}, \ and\
  \bibinfo {author} {\bibfnamefont {N.}~\bibnamefont {Mingo}},\ }\href
  {\doibase 10.1063/1.4850995} {\bibfield  {journal} {\bibinfo  {journal}
  {Applied Physics Letters}\ }\textbf {\bibinfo {volume} {103}},\ \bibinfo
  {pages} {253103} (\bibinfo {year} {2013}{\natexlab{a}})}\BibitemShut
  {NoStop}%
\bibitem [{\citenamefont {Kaasbjerg}\ \emph {et~al.}(2014)\citenamefont
  {Kaasbjerg}, \citenamefont {Bhargavi},\ and\ \citenamefont
  {Kubakaddi}}]{kaasbjerg_hot-electron_2014}%
  \BibitemOpen
  \bibfield  {author} {\bibinfo {author} {\bibfnamefont {K.}~\bibnamefont
  {Kaasbjerg}}, \bibinfo {author} {\bibfnamefont {K.~S.}\ \bibnamefont
  {Bhargavi}}, \ and\ \bibinfo {author} {\bibfnamefont {S.~S.}\ \bibnamefont
  {Kubakaddi}},\ }\href {\doibase 10.1103/PhysRevB.90.165436} {\bibfield
  {journal} {\bibinfo  {journal} {Physical Review B}\ }\textbf {\bibinfo
  {volume} {90}},\ \bibinfo {pages} {165436} (\bibinfo {year}
  {2014})}\BibitemShut {NoStop}%
\bibitem [{\citenamefont {Ge}\ \emph {et~al.}(2014)\citenamefont {Ge},
  \citenamefont {Wan}, \citenamefont {Feng}, \citenamefont {Xiao},\ and\
  \citenamefont {Yao}}]{ge_effect_2014}%
  \BibitemOpen
  \bibfield  {author} {\bibinfo {author} {\bibfnamefont {Y.}~\bibnamefont
  {Ge}}, \bibinfo {author} {\bibfnamefont {W.}~\bibnamefont {Wan}}, \bibinfo
  {author} {\bibfnamefont {W.}~\bibnamefont {Feng}}, \bibinfo {author}
  {\bibfnamefont {D.}~\bibnamefont {Xiao}}, \ and\ \bibinfo {author}
  {\bibfnamefont {Y.}~\bibnamefont {Yao}},\ }\href {\doibase
  10.1103/PhysRevB.90.035414} {\bibfield  {journal} {\bibinfo  {journal}
  {Physical Review B}\ }\textbf {\bibinfo {volume} {90}},\ \bibinfo {pages}
  {035414} (\bibinfo {year} {2014})}\BibitemShut {NoStop}%
\bibitem [{\citenamefont {Tongay}\ \emph {et~al.}(2012)\citenamefont {Tongay},
  \citenamefont {Zhou}, \citenamefont {Ataca}, \citenamefont {Lo},
  \citenamefont {Matthews}, \citenamefont {Li}, \citenamefont {Grossman},\ and\
  \citenamefont {Wu}}]{Tongay2012}%
  \BibitemOpen
  \bibfield  {author} {\bibinfo {author} {\bibfnamefont {S.}~\bibnamefont
  {Tongay}}, \bibinfo {author} {\bibfnamefont {J.}~\bibnamefont {Zhou}},
  \bibinfo {author} {\bibfnamefont {C.}~\bibnamefont {Ataca}}, \bibinfo
  {author} {\bibfnamefont {K.}~\bibnamefont {Lo}}, \bibinfo {author}
  {\bibfnamefont {T.~S.}\ \bibnamefont {Matthews}}, \bibinfo {author}
  {\bibfnamefont {J.}~\bibnamefont {Li}}, \bibinfo {author} {\bibfnamefont
  {J.~C.}\ \bibnamefont {Grossman}}, \ and\ \bibinfo {author} {\bibfnamefont
  {J.}~\bibnamefont {Wu}},\ }\href {\doibase 10.1021/nl302584w} {\bibfield
  {journal} {\bibinfo  {journal} {Nano Letters}\ }\textbf {\bibinfo {volume}
  {12}},\ \bibinfo {pages} {5576} (\bibinfo {year} {2012})},\ \bibinfo {note}
  {pMID: 23098085},\ \Eprint
  {http://arxiv.org/abs/http://dx.doi.org/10.1021/nl302584w}
  {http://dx.doi.org/10.1021/nl302584w} \BibitemShut {NoStop}%
\bibitem [{\citenamefont {Li}\ \emph {et~al.}(2014)\citenamefont {Li},
  \citenamefont {Chernikov}, \citenamefont {Zhang}, \citenamefont {Rigosi},
  \citenamefont {Hill}, \citenamefont {van~der Zande}, \citenamefont {Chenet},
  \citenamefont {Shih}, \citenamefont {Hone},\ and\ \citenamefont
  {Heinz}}]{li_measurement_2014}%
  \BibitemOpen
  \bibfield  {author} {\bibinfo {author} {\bibfnamefont {Y.}~\bibnamefont
  {Li}}, \bibinfo {author} {\bibfnamefont {A.}~\bibnamefont {Chernikov}},
  \bibinfo {author} {\bibfnamefont {X.}~\bibnamefont {Zhang}}, \bibinfo
  {author} {\bibfnamefont {A.}~\bibnamefont {Rigosi}}, \bibinfo {author}
  {\bibfnamefont {H.~M.}\ \bibnamefont {Hill}}, \bibinfo {author}
  {\bibfnamefont {A.~M.}\ \bibnamefont {van~der Zande}}, \bibinfo {author}
  {\bibfnamefont {D.~A.}\ \bibnamefont {Chenet}}, \bibinfo {author}
  {\bibfnamefont {E.-M.}\ \bibnamefont {Shih}}, \bibinfo {author}
  {\bibfnamefont {J.}~\bibnamefont {Hone}}, \ and\ \bibinfo {author}
  {\bibfnamefont {T.~F.}\ \bibnamefont {Heinz}},\ }\href {\doibase
  10.1103/PhysRevB.90.205422} {\bibfield  {journal} {\bibinfo  {journal}
  {Physical Review B}\ }\textbf {\bibinfo {volume} {90}},\ \bibinfo {pages}
  {205422} (\bibinfo {year} {2014})}\BibitemShut {NoStop}%
\bibitem [{\citenamefont {Latzke}\ \emph {et~al.}(2015)\citenamefont {Latzke},
  \citenamefont {Zhang}, \citenamefont {Suslu}, \citenamefont {Chang},
  \citenamefont {Lin}, \citenamefont {Jeng}, \citenamefont {Tongay},
  \citenamefont {Wu}, \citenamefont {Bansil},\ and\ \citenamefont
  {Lanzara}}]{latzke_electronic_2015}%
  \BibitemOpen
  \bibfield  {author} {\bibinfo {author} {\bibfnamefont {D.~W.}\ \bibnamefont
  {Latzke}}, \bibinfo {author} {\bibfnamefont {W.}~\bibnamefont {Zhang}},
  \bibinfo {author} {\bibfnamefont {A.}~\bibnamefont {Suslu}}, \bibinfo
  {author} {\bibfnamefont {T.-R.}\ \bibnamefont {Chang}}, \bibinfo {author}
  {\bibfnamefont {H.}~\bibnamefont {Lin}}, \bibinfo {author} {\bibfnamefont
  {H.-T.}\ \bibnamefont {Jeng}}, \bibinfo {author} {\bibfnamefont
  {S.}~\bibnamefont {Tongay}}, \bibinfo {author} {\bibfnamefont
  {J.}~\bibnamefont {Wu}}, \bibinfo {author} {\bibfnamefont {A.}~\bibnamefont
  {Bansil}}, \ and\ \bibinfo {author} {\bibfnamefont {A.}~\bibnamefont
  {Lanzara}},\ }\href {\doibase 10.1103/PhysRevB.91.235202} {\bibfield
  {journal} {\bibinfo  {journal} {Physical Review B}\ }\textbf {\bibinfo
  {volume} {91}},\ \bibinfo {pages} {235202} (\bibinfo {year}
  {2015})}\BibitemShut {NoStop}%
\bibitem [{\citenamefont {Carvalho}\ \emph {et~al.}(2015)\citenamefont
  {Carvalho}, \citenamefont {Malard}, \citenamefont {Alves}, \citenamefont
  {Fantini},\ and\ \citenamefont {Pimenta}}]{carvalho_symmetry-dependent_2015}%
  \BibitemOpen
  \bibfield  {author} {\bibinfo {author} {\bibfnamefont {B.~R.}\ \bibnamefont
  {Carvalho}}, \bibinfo {author} {\bibfnamefont {L.~M.}\ \bibnamefont
  {Malard}}, \bibinfo {author} {\bibfnamefont {J.~M.}\ \bibnamefont {Alves}},
  \bibinfo {author} {\bibfnamefont {C.}~\bibnamefont {Fantini}}, \ and\
  \bibinfo {author} {\bibfnamefont {M.~A.}\ \bibnamefont {Pimenta}},\ }\href
  {\doibase 10.1103/PhysRevLett.114.136403} {\bibfield  {journal} {\bibinfo
  {journal} {Physical Review Letters}\ }\textbf {\bibinfo {volume} {114}},\
  \bibinfo {pages} {136403} (\bibinfo {year} {2015})}\BibitemShut {NoStop}%
\bibitem [{\citenamefont {Antonius}\ \emph {et~al.}(2014)\citenamefont
  {Antonius}, \citenamefont {Ponc\'{e}}, \citenamefont {Boulanger},
  \citenamefont {Cot\'{e}},\ and\ \citenamefont
  {Gonze}}]{antonius_many-body_2014}%
  \BibitemOpen
  \bibfield  {author} {\bibinfo {author} {\bibfnamefont {G.}~\bibnamefont
  {Antonius}}, \bibinfo {author} {\bibfnamefont {S.}~\bibnamefont {Ponc\'{e}}},
  \bibinfo {author} {\bibfnamefont {P.}~\bibnamefont {Boulanger}}, \bibinfo
  {author} {\bibfnamefont {M.}~\bibnamefont {Cot\'{e}}}, \ and\ \bibinfo
  {author} {\bibfnamefont {X.}~\bibnamefont {Gonze}},\ }\href {\doibase
  10.1103/PhysRevLett.112.215501} {\bibfield  {journal} {\bibinfo  {journal}
  {Physical Review Letters}\ }\textbf {\bibinfo {volume} {112}},\ \bibinfo
  {pages} {215501} (\bibinfo {year} {2014})}\BibitemShut {NoStop}%
\bibitem [{\citenamefont {Marini}\ \emph {et~al.}(2015)\citenamefont {Marini},
  \citenamefont {Ponc\'e},\ and\ \citenamefont {Gonze}}]{Marini2015}%
  \BibitemOpen
  \bibfield  {author} {\bibinfo {author} {\bibfnamefont {A.}~\bibnamefont
  {Marini}}, \bibinfo {author} {\bibfnamefont {S.}~\bibnamefont {Ponc\'e}}, \
  and\ \bibinfo {author} {\bibfnamefont {X.}~\bibnamefont {Gonze}},\ }\href
  {\doibase 10.1103/PhysRevB.91.224310} {\bibfield  {journal} {\bibinfo
  {journal} {Phys. Rev. B}\ }\textbf {\bibinfo {volume} {91}},\ \bibinfo
  {pages} {224310} (\bibinfo {year} {2015})}\BibitemShut {NoStop}%
\bibitem [{\citenamefont {Marini}(2008)}]{marini_textitab_2008}%
  \BibitemOpen
  \bibfield  {author} {\bibinfo {author} {\bibfnamefont {A.}~\bibnamefont
  {Marini}},\ }\href {\doibase 10.1103/PhysRevLett.101.106405} {\bibfield
  {journal} {\bibinfo  {journal} {Physical Review Letters}\ }\textbf {\bibinfo
  {volume} {101}},\ \bibinfo {pages} {106405} (\bibinfo {year}
  {2008})}\BibitemShut {NoStop}%
\bibitem [{\citenamefont {Giannozzi}\ \emph {et~al.}(2009)\citenamefont
  {Giannozzi}, \citenamefont {Baroni}, \citenamefont {Bonini}, \citenamefont
  {Calandra}, \citenamefont {Car}, \citenamefont {Cavazzoni}, \citenamefont
  {Ceresoli}, \citenamefont {Chiarotti}, \citenamefont {Cococcioni},
  \citenamefont {Dabo}, \citenamefont {Corso}, \citenamefont {Gironcoli},
  \citenamefont {Fabris}, \citenamefont {Fratesi}, \citenamefont {Gebauer},
  \citenamefont {Gerstmann}, \citenamefont {Gougoussis}, \citenamefont
  {Kokalj}, \citenamefont {Lazzeri}, \citenamefont {Martin-Samos},
  \citenamefont {Marzari}, \citenamefont {Mauri}, \citenamefont {Mazzarello},
  \citenamefont {Paolini}, \citenamefont {Pasquarello}, \citenamefont
  {Paulatto}, \citenamefont {Sbraccia}, \citenamefont {Scandolo}, \citenamefont
  {Sclauzero}, \citenamefont {Seitsonen}, \citenamefont {Smogunov},
  \citenamefont {Umari},\ and\ \citenamefont
  {Wentzcovitch}}]{giannozzi_quantum_2009}%
  \BibitemOpen
  \bibfield  {author} {\bibinfo {author} {\bibfnamefont {P.}~\bibnamefont
  {Giannozzi}}, \bibinfo {author} {\bibfnamefont {S.}~\bibnamefont {Baroni}},
  \bibinfo {author} {\bibfnamefont {N.}~\bibnamefont {Bonini}}, \bibinfo
  {author} {\bibfnamefont {M.}~\bibnamefont {Calandra}}, \bibinfo {author}
  {\bibfnamefont {R.}~\bibnamefont {Car}}, \bibinfo {author} {\bibfnamefont
  {C.}~\bibnamefont {Cavazzoni}}, \bibinfo {author} {\bibfnamefont
  {D.}~\bibnamefont {Ceresoli}}, \bibinfo {author} {\bibfnamefont {G.~L.}\
  \bibnamefont {Chiarotti}}, \bibinfo {author} {\bibfnamefont {M.}~\bibnamefont
  {Cococcioni}}, \bibinfo {author} {\bibfnamefont {I.}~\bibnamefont {Dabo}},
  \bibinfo {author} {\bibfnamefont {A.~D.}\ \bibnamefont {Corso}}, \bibinfo
  {author} {\bibfnamefont {S.~d.}\ \bibnamefont {Gironcoli}}, \bibinfo {author}
  {\bibfnamefont {S.}~\bibnamefont {Fabris}}, \bibinfo {author} {\bibfnamefont
  {G.}~\bibnamefont {Fratesi}}, \bibinfo {author} {\bibfnamefont
  {R.}~\bibnamefont {Gebauer}}, \bibinfo {author} {\bibfnamefont
  {U.}~\bibnamefont {Gerstmann}}, \bibinfo {author} {\bibfnamefont
  {C.}~\bibnamefont {Gougoussis}}, \bibinfo {author} {\bibfnamefont
  {A.}~\bibnamefont {Kokalj}}, \bibinfo {author} {\bibfnamefont
  {M.}~\bibnamefont {Lazzeri}}, \bibinfo {author} {\bibfnamefont
  {L.}~\bibnamefont {Martin-Samos}}, \bibinfo {author} {\bibfnamefont
  {N.}~\bibnamefont {Marzari}}, \bibinfo {author} {\bibfnamefont
  {F.}~\bibnamefont {Mauri}}, \bibinfo {author} {\bibfnamefont
  {R.}~\bibnamefont {Mazzarello}}, \bibinfo {author} {\bibfnamefont
  {S.}~\bibnamefont {Paolini}}, \bibinfo {author} {\bibfnamefont
  {A.}~\bibnamefont {Pasquarello}}, \bibinfo {author} {\bibfnamefont
  {L.}~\bibnamefont {Paulatto}}, \bibinfo {author} {\bibfnamefont
  {C.}~\bibnamefont {Sbraccia}}, \bibinfo {author} {\bibfnamefont
  {S.}~\bibnamefont {Scandolo}}, \bibinfo {author} {\bibfnamefont
  {G.}~\bibnamefont {Sclauzero}}, \bibinfo {author} {\bibfnamefont {A.~P.}\
  \bibnamefont {Seitsonen}}, \bibinfo {author} {\bibfnamefont {A.}~\bibnamefont
  {Smogunov}}, \bibinfo {author} {\bibfnamefont {P.}~\bibnamefont {Umari}}, \
  and\ \bibinfo {author} {\bibfnamefont {R.~M.}\ \bibnamefont {Wentzcovitch}},\
  }\href {\doibase 10.1088/0953-8984/21/39/395502} {\bibfield  {journal}
  {\bibinfo  {journal} {Journal of Physics: Condensed Matter}\ }\textbf
  {\bibinfo {volume} {21}},\ \bibinfo {pages} {395502} (\bibinfo {year}
  {2009})}\BibitemShut {NoStop}%
\bibitem [{\citenamefont {Cannuccia}\ and\ \citenamefont
  {Marini}(2011)}]{cannuccia_effect_2011}%
  \BibitemOpen
  \bibfield  {author} {\bibinfo {author} {\bibfnamefont {E.}~\bibnamefont
  {Cannuccia}}\ and\ \bibinfo {author} {\bibfnamefont {A.}~\bibnamefont
  {Marini}},\ }\href {\doibase 10.1103/PhysRevLett.107.255501} {\bibfield
  {journal} {\bibinfo  {journal} {Physical Review Letters}\ }\textbf {\bibinfo
  {volume} {107}},\ \bibinfo {pages} {255501} (\bibinfo {year}
  {2011})}\BibitemShut {NoStop}%
\bibitem [{\citenamefont {Verdi}\ and\ \citenamefont
  {Giustino}(2015)}]{Verdi2015}%
  \BibitemOpen
  \bibfield  {author} {\bibinfo {author} {\bibfnamefont {C.}~\bibnamefont
  {Verdi}}\ and\ \bibinfo {author} {\bibfnamefont {F.}~\bibnamefont
  {Giustino}},\ }\href {\doibase 10.1103/PhysRevLett.115.176401} {\bibfield
  {journal} {\bibinfo  {journal} {Phys. Rev. Lett.}\ }\textbf {\bibinfo
  {volume} {115}},\ \bibinfo {pages} {176401} (\bibinfo {year}
  {2015})}\BibitemShut {NoStop}%
\bibitem [{\citenamefont {Marini}\ \emph {et~al.}(2009)\citenamefont {Marini},
  \citenamefont {Hogan}, \citenamefont {Gr\"uning},\ and\ \citenamefont
  {Varsano}}]{marini_yambo:_2009}%
  \BibitemOpen
  \bibfield  {author} {\bibinfo {author} {\bibfnamefont {A.}~\bibnamefont
  {Marini}}, \bibinfo {author} {\bibfnamefont {C.}~\bibnamefont {Hogan}},
  \bibinfo {author} {\bibfnamefont {M.}~\bibnamefont {Gr\"uning}}, \ and\
  \bibinfo {author} {\bibfnamefont {D.}~\bibnamefont {Varsano}},\ }\href
  {\doibase 10.1016/j.cpc.2009.02.003} {\bibfield  {journal} {\bibinfo
  {journal} {Computer Physics Communications}\ }\textbf {\bibinfo {volume}
  {180}},\ \bibinfo {pages} {1392} (\bibinfo {year} {2009})}\BibitemShut
  {NoStop}%
\bibitem [{\citenamefont {Ponc\'e}\ \emph {et~al.}(2015)\citenamefont
  {Ponc\'e}, \citenamefont {Gillet}, \citenamefont {Laflamme~Janssen},
  \citenamefont {Marini}, \citenamefont {Verstraete},\ and\ \citenamefont
  {Gonze}}]{Ponce2015}%
  \BibitemOpen
  \bibfield  {author} {\bibinfo {author} {\bibfnamefont {S.}~\bibnamefont
  {Ponc\'e}}, \bibinfo {author} {\bibfnamefont {Y.}~\bibnamefont {Gillet}},
  \bibinfo {author} {\bibfnamefont {J.}~\bibnamefont {Laflamme~Janssen}},
  \bibinfo {author} {\bibfnamefont {A.}~\bibnamefont {Marini}}, \bibinfo
  {author} {\bibfnamefont {M.}~\bibnamefont {Verstraete}}, \ and\ \bibinfo
  {author} {\bibfnamefont {X.}~\bibnamefont {Gonze}},\ }\href {\doibase
  http://dx.doi.org/10.1063/1.4927081} {\bibfield  {journal} {\bibinfo
  {journal} {The Journal of Chemical Physics}\ }\textbf {\bibinfo {volume}
  {143}},\ \bibinfo {eid} {102813} (\bibinfo {year} {2015}),\
  http://dx.doi.org/10.1063/1.4927081}\BibitemShut {NoStop}%
\bibitem [{\citenamefont {Stone}\ \emph {et~al.}(2006)\citenamefont {Stone},
  \citenamefont {Zaliznyak}, \citenamefont {Hong}, \citenamefont {Broholm},\
  and\ \citenamefont {Reich}}]{stone_quasiparticle_2006}%
  \BibitemOpen
  \bibfield  {author} {\bibinfo {author} {\bibfnamefont {M.~B.}\ \bibnamefont
  {Stone}}, \bibinfo {author} {\bibfnamefont {I.~A.}\ \bibnamefont
  {Zaliznyak}}, \bibinfo {author} {\bibfnamefont {T.}~\bibnamefont {Hong}},
  \bibinfo {author} {\bibfnamefont {C.~L.}\ \bibnamefont {Broholm}}, \ and\
  \bibinfo {author} {\bibfnamefont {D.~H.}\ \bibnamefont {Reich}},\ }\href
  {\doibase 10.1038/nature04593} {\bibfield  {journal} {\bibinfo  {journal}
  {Nature}\ }\textbf {\bibinfo {volume} {440}},\ \bibinfo {pages} {187}
  (\bibinfo {year} {2006})}\BibitemShut {NoStop}%
\bibitem [{Note1()}]{Note1}%
  \BibitemOpen
  \bibinfo {note} {By spin mixing we mean that an electron in the VBM and with
  momentum $K$ and spin up or mostly up can relax into a state at $K$ with spin
  down or mostly down.}\BibitemShut {Stop}%
\bibitem [{\citenamefont {Li}\ \emph {et~al.}(2013{\natexlab{b}})\citenamefont
  {Li}, \citenamefont {Mullen}, \citenamefont {Jin}, \citenamefont {Borysenko},
  \citenamefont {Buongiorno~Nardelli},\ and\ \citenamefont
  {Kim}}]{li_intrinsic_2013}%
  \BibitemOpen
  \bibfield  {author} {\bibinfo {author} {\bibfnamefont {X.}~\bibnamefont
  {Li}}, \bibinfo {author} {\bibfnamefont {J.~T.}\ \bibnamefont {Mullen}},
  \bibinfo {author} {\bibfnamefont {Z.}~\bibnamefont {Jin}}, \bibinfo {author}
  {\bibfnamefont {K.~M.}\ \bibnamefont {Borysenko}}, \bibinfo {author}
  {\bibfnamefont {M.}~\bibnamefont {Buongiorno~Nardelli}}, \ and\ \bibinfo
  {author} {\bibfnamefont {K.~W.}\ \bibnamefont {Kim}},\ }\href {\doibase
  10.1103/PhysRevB.87.115418} {\bibfield  {journal} {\bibinfo  {journal}
  {Physical Review B}\ }\textbf {\bibinfo {volume} {87}},\ \bibinfo {pages}
  {115418} (\bibinfo {year} {2013}{\natexlab{b}})}\BibitemShut {NoStop}%
\bibitem [{\citenamefont {Molina-S\'anchez}\ and\ \citenamefont
  {Wirtz}(2011)}]{molina-sanchez_phonons_2011}%
  \BibitemOpen
  \bibfield  {author} {\bibinfo {author} {\bibfnamefont {A.}~\bibnamefont
  {Molina-S\'anchez}}\ and\ \bibinfo {author} {\bibfnamefont {L.}~\bibnamefont
  {Wirtz}},\ }\href {\doibase 10.1103/PhysRevB.84.155413} {\bibfield  {journal}
  {\bibinfo  {journal} {Physical Review B}\ }\textbf {\bibinfo {volume} {84}},\
  \bibinfo {pages} {155413} (\bibinfo {year} {2011})}\BibitemShut {NoStop}%
\bibitem [{\citenamefont {Cannuccia}\ and\ \citenamefont
  {Marini}(2012)}]{cannuccia_zero_2012}%
  \BibitemOpen
  \bibfield  {author} {\bibinfo {author} {\bibfnamefont {E.}~\bibnamefont
  {Cannuccia}}\ and\ \bibinfo {author} {\bibfnamefont {A.}~\bibnamefont
  {Marini}},\ }\href {\doibase 10.1140/epjb/e2012-30105-4} {\bibfield
  {journal} {\bibinfo  {journal} {The European Physical Journal B}\ }\textbf
  {\bibinfo {volume} {85}},\ \bibinfo {pages} {1} (\bibinfo {year}
  {2012})}\BibitemShut {NoStop}%
\bibitem [{\citenamefont {Monserrat}\ and\ \citenamefont
  {Needs}(2014)}]{monserrat_comparing_2014}%
  \BibitemOpen
  \bibfield  {author} {\bibinfo {author} {\bibfnamefont {B.}~\bibnamefont
  {Monserrat}}\ and\ \bibinfo {author} {\bibfnamefont {R.~J.}\ \bibnamefont
  {Needs}},\ }\href {\doibase 10.1103/PhysRevB.89.214304} {\bibfield  {journal}
  {\bibinfo  {journal} {Physical Review B}\ }\textbf {\bibinfo {volume} {89}},\
  \bibinfo {pages} {214304} (\bibinfo {year} {2014})}\BibitemShut {NoStop}%
\bibitem [{\citenamefont {Gillet}\ \emph {et~al.}(2013)\citenamefont {Gillet},
  \citenamefont {Giantomassi},\ and\ \citenamefont
  {Gonze}}]{gillet_first-principles_2013}%
  \BibitemOpen
  \bibfield  {author} {\bibinfo {author} {\bibfnamefont {Y.}~\bibnamefont
  {Gillet}}, \bibinfo {author} {\bibfnamefont {M.}~\bibnamefont {Giantomassi}},
  \ and\ \bibinfo {author} {\bibfnamefont {X.}~\bibnamefont {Gonze}},\ }\href
  {\doibase 10.1103/PhysRevB.88.094305} {\bibfield  {journal} {\bibinfo
  {journal} {Physical Review B}\ }\textbf {\bibinfo {volume} {88}},\ \bibinfo
  {pages} {094305} (\bibinfo {year} {2013})}\BibitemShut {NoStop}%
\bibitem [{\citenamefont {Korn}\ \emph {et~al.}(2011)\citenamefont {Korn},
  \citenamefont {Heydrich}, \citenamefont {Hirmer}, \citenamefont
  {Schmutzler},\ and\ \citenamefont
  {Sch\"{u}ller}}]{korn_low-temperature_2011}%
  \BibitemOpen
  \bibfield  {author} {\bibinfo {author} {\bibfnamefont {T.}~\bibnamefont
  {Korn}}, \bibinfo {author} {\bibfnamefont {S.}~\bibnamefont {Heydrich}},
  \bibinfo {author} {\bibfnamefont {M.}~\bibnamefont {Hirmer}}, \bibinfo
  {author} {\bibfnamefont {J.}~\bibnamefont {Schmutzler}}, \ and\ \bibinfo
  {author} {\bibfnamefont {C.}~\bibnamefont {Sch\"{u}ller}},\ }\href {\doibase
  10.1063/1.3636402} {\bibfield  {journal} {\bibinfo  {journal} {Applied
  Physics Letters}\ }\textbf {\bibinfo {volume} {99}},\ \bibinfo {pages}
  {102109} (\bibinfo {year} {2011})}\BibitemShut {NoStop}%
\bibitem [{\citenamefont {Rohlfing}\ and\ \citenamefont
  {Louie}(2000)}]{rohlfing_electron-hole_2000}%
  \BibitemOpen
  \bibfield  {author} {\bibinfo {author} {\bibfnamefont {M.}~\bibnamefont
  {Rohlfing}}\ and\ \bibinfo {author} {\bibfnamefont {S.~G.}\ \bibnamefont
  {Louie}},\ }\href {\doibase 10.1103/PhysRevB.62.4927} {\bibfield  {journal}
  {\bibinfo  {journal} {Physical Review B}\ }\textbf {\bibinfo {volume} {62}},\
  \bibinfo {pages} {4927} (\bibinfo {year} {2000})}\BibitemShut {NoStop}%
\bibitem [{\citenamefont {Onida}\ \emph {et~al.}(2002)\citenamefont {Onida},
  \citenamefont {Reining},\ and\ \citenamefont {Rubio}}]{Onida2002}%
  \BibitemOpen
  \bibfield  {author} {\bibinfo {author} {\bibfnamefont {G.}~\bibnamefont
  {Onida}}, \bibinfo {author} {\bibfnamefont {L.}~\bibnamefont {Reining}}, \
  and\ \bibinfo {author} {\bibfnamefont {A.}~\bibnamefont {Rubio}},\ }\href
  {\doibase 10.1103/RevModPhys.74.601} {\bibfield  {journal} {\bibinfo
  {journal} {Rev. Mod. Phys.}\ }\textbf {\bibinfo {volume} {74}},\ \bibinfo
  {pages} {601} (\bibinfo {year} {2002})}\BibitemShut {NoStop}%
\bibitem [{\citenamefont {Shi}\ \emph {et~al.}(2013)\citenamefont {Shi},
  \citenamefont {Pan}, \citenamefont {Zhang},\ and\ \citenamefont
  {Yakobson}}]{shi_quasiparticle_2013}%
  \BibitemOpen
  \bibfield  {author} {\bibinfo {author} {\bibfnamefont {H.}~\bibnamefont
  {Shi}}, \bibinfo {author} {\bibfnamefont {H.}~\bibnamefont {Pan}}, \bibinfo
  {author} {\bibfnamefont {Y.-W.}\ \bibnamefont {Zhang}}, \ and\ \bibinfo
  {author} {\bibfnamefont {B.~I.}\ \bibnamefont {Yakobson}},\ }\href {\doibase
  10.1103/PhysRevB.87.155304} {\bibfield  {journal} {\bibinfo  {journal}
  {Physical Review B}\ }\textbf {\bibinfo {volume} {87}},\ \bibinfo {pages}
  {155304} (\bibinfo {year} {2013})}\BibitemShut {NoStop}%
\end{thebibliography}
\end{document}